\documentclass[12pt,preprint]{aastex}

\slugcomment{draft of \today}
\shorttitle{RM toward High Galactic Latitudes}
\shortauthors{Akahori et al.}

\begin{document}

\title{Simulated Faraday Rotation Measures toward High Galactic Latitudes}
\author{Takuya Akahori$^{1,2}$, Dongsu Ryu$^3$\altaffilmark{,4}, Jongsoo Kim$^1$, and B. M. Gaensler$^2$}
\affil{$^1$Korea Astronomy and Space Science Institute, Daedeokdaero 776, Yuseong-Gu, Daejeon 305-348, Korea: jskim@kasi.re.kr\\
$^2$Sydney Institute for Astronomy, School of Physics, The University of Sydney, NSW 2006, Australia: akahori@physics.usyd.edu.au, bryan.gaensler@sydney.edu.au\\
$^3$Department of Astronomy and Space Science, Chungnam National University, Daejeon 305-764, Korea: ryu@canopus.cnu.ac.kr}
\altaffiltext{4}{Corresponding Author}

\begin{abstract}

We study the Faraday rotation measure (RM) due to the Galactic magnetic field (GMF) toward high Galactic latitudes. The RM arises from the global, regular component as well as from the turbulent, random component of the GMF. We model the former based on observations and the latter using the data of magnetohydrodynamic turbulence simulations. For a large number of different GMF models, we produce mock RM maps around the Galactic poles and calculate various statistical quantities with the RM maps. We find that the observed medians of RMs toward the north and south Galactic poles, $\sim 0.0\pm 0.5~{\rm rad~m^{-2}}$ and $\sim +6.3\pm 0.5~{\rm rad~m^{-2}}$, are difficult to explain with any of our many alternate GMF models. The standard deviation of observed RMs, $\sim 9~{\rm rad~m^{-2}}$, is clearly larger than that of simulated RMs. The second-order structure function of observed RMs is substantially larger than that of simulated RMs, especially at small angular scales. We discuss other possible contributions to RM toward high Galactic latitudes. Besides observational errors and the intrinsic RM of background radio sources against which RM is observed, we suggest that the RM due to the intergalactic magnetic field may account for a substantial fraction of the observed RM. Finally we note that reproducing the observed medians may require additional components or/and structures of the GMF that are not present in our models.
\end{abstract}


\section{Introduction}
\label{section1}

{\it Cosmic magnetism}, the origin and nature of magnetic fields in our universe, is one of outstanding problems of modern astrophysics \citep[see][]{gbf04}. Exploration of the Galactic magnetic field (GMF) and the intergalactic magnetic field (IGMF) is listed as one of key science projects for the Square Kilometer Array (SKA), and is one of the important science projects for Jansky Very Large Array (JVLA), Murchison Widefield Array (MWA), the Low Frequency Array (LOFAR), the Australian SKA Pathfinder (ASKAP), and the South African Karoo Array Telescope (MeerKAT) \citep[see, e.g.,][]{cr04,beck09a,kra09,glt10}. Measuring Faraday rotation, the rotation of the plane of linearly-polarized radio emission due to the birefringence of magneto-ionic medium, is a powerful method to observe the GMF as well as the IGMF. Up to now, there have been a number of studies of Faraday rotation measure (RM) to elucidate the structure and statistical properties of the GMF and the IGMF.

For the IGMF, there have been a number of observations of RM in clusters of galaxies \citep[see][for a review]{ct02}. Observations have been also extended to the outside of clusters, toward the cosmic web, and extragalactic contributions of RM due to the IGMF or others have been discussed \citep[see, e.g.,][]{xkhd06,kro08,ber12,ham12}. There are several theoretical works for RM in filaments of galaxies \citep{rkb98,dcst05,cr09,sndbd10,ar10,ar11}. For instance, \citet{ar10} investigated RM in filaments using a model IGMF based on a turbulent dynamo \citep{rkcd08}. They found that the root-mean-square (rms) value of RMs through a single filament is expected to be $\sim 1~{\rm rad~m^{-2}}$ in the local universe. \citet{ar11} extended this work by using the redshift distribution of polarized background radio sources against which RM is observed. They found that the rms value of RMs through filaments up to redshift $\sim 5$ would be $\sim {\rm several}~{\rm rad~m^{-2}}$. They also found that the second-order structure function (SF) has a nearly-flat profile in angular separations of $\gtrsim 0.2^\circ$, meaning that RMs through filaments decorrelate on angles less than $\sim 0.2^\circ$.

Recently, the RMs toward the North and South Galactic poles (NGP and SGP, respectively) have been investigated to study the GMF as well as the IGMF. \citet{tss09} studied the RM data from the NRAO VLA Sky Survey (NVSS), and estimated non-zero vertical strengths of the GMF, about $-0.14\pm 0.02~\mu{\rm G}$ and $+0.3\pm 0.03~\mu{\rm G}$ toward the NGP and SGP, respectively. \citet{mao10} used RM data from the Westerbork Radio Synthesis Telescope (WSRT) and the Australia Telescope Compact Array (ATCA), and found that the median value of RMs toward the SGP is $+6.3\pm 0.5~ {\rm rad~m^{-2}}$ (corresponding to a vertical GMF strength of $+0.31\pm 0.02~\mu{\rm G}$), while that toward the NGP is $0.0\pm 0.5~{\rm rad~m^{-2}}$ ($+0.00\pm 0.02~\mu{\rm G}$). The standard deviations of RMs were $\simeq 9.2~{\rm rad~m^{-2}}$ and $8.8~{\rm rad~m^{-2}}$ toward the NGP and SGP, respectively. \citet{mao10} also put an upper limit of $\sim 1~\mu{\rm G}$ on the strength of random magnetic field at high Galactic latitudes. \citet{sts11} examined the NVSS data in detail, and found that the second-order SFs of RM at angular separations of $\gtrsim 1^\circ$ have a value $\sim 100-200~{\rm rad^2~m^{-4}}$ toward the NGP and $\sim 300-400~{\rm rad^2~m^{-4}}$ toward the SGP. In addition, based on the latitude dependence of RM, \citet{sch10} examined the Galactic and extragalactic contributions to RM in the NVSS data. He estimated that the Galactic contribution is $\bar{\sigma}_{\rm RM,MW}\sim 6.8 \pm 0.1 (8.4 \pm 0.1)~{\rm rad~m^{-2}}$ and the extragalactic contribution (including those intrinsic to the polarized background radio sources and due to the IGMF) is $\bar{\sigma}_{\rm RM,EG}\sim 6.5 \pm 0.1 (5.9 \pm 0.2)~{\rm rad~m^{-2}}$ for the northern (southern) hemisphere.

The separation of the Galactic and extragalactic contributions in observed RMs, however, is not trivial. It requires good understandings of the GMF and its contribution to observed RMs. There have been a number of theoretical works to model the Galactic RM \citep[e.g.,][]{sun08,wae09,sr09,jaf10,van11,ptkn11,mao12,jf12} \citep[see also works on cosmic-ray propagation:][]{ps03,tt05,gkss10,ts10}. The Galactic RM arises from the global, regular component as well as from the turbulent, random component of the GMF. While the regular component has been modeled with analytic fitting formulae based on observations, the random component has been modeled using power-law spectra with random phases in Fourier space. For instance, \citet{sr09} used the publicly-available HAMMURABI code \citep{wae09}, and adopted a Kolmogorov-like power spectrum with average amplitude $3~\mu{\rm G}$ in a box of 10 pc size. They found that at Galactic latitudes $|b|\sim 70^\circ$, the second-order SF has a magnitude of up to a few $\times 100\ {\rm rad^2~m^{-4}}$ at angular scales of $\ga 10'$.

Previous studies have successfully reproduced the observed properties of the Galactic RM as well as those of the radio continuum emission toward low and mid-Galactic latitudes \citep[e.g.,][]{sun08}. There is, however, a lack of studies that can be compared with recent observations of RMs toward the Galactic poles. In this paper, we simulate the Galactic RM toward high Galactic latitudes and investigate its statistical properties. While we model the regular component of the GMF based on a number of observations, we use the data of three-dimensional magnetohydrodynamic (MHD) turbulence simulations to model the random component. We obtain simulated maps of the Galactic RM toward the Galactic poles in a field-of-view (FOV) of $900~{\rm deg^2}$ to compare with recent observations and in a $200~{\rm deg^2}$ FOV to predict future SKA observations. We estimate how much the GMF can contribute to observed RMs and discuss how the statistics of simulated RMs compare with those of observed RMs. The rest of the paper is organized as follows. In Section 2, we describe our models. The results are shown in Section 3. Discussion and summary follow in Sections 4 and 5, respectively.

\section{Models}
\label{section2}

\subsection{Regular Components}
\label{section2.1}

\begin{figure}[tp]
\figurenum{1}
\epsscale{1.0}
\plotone{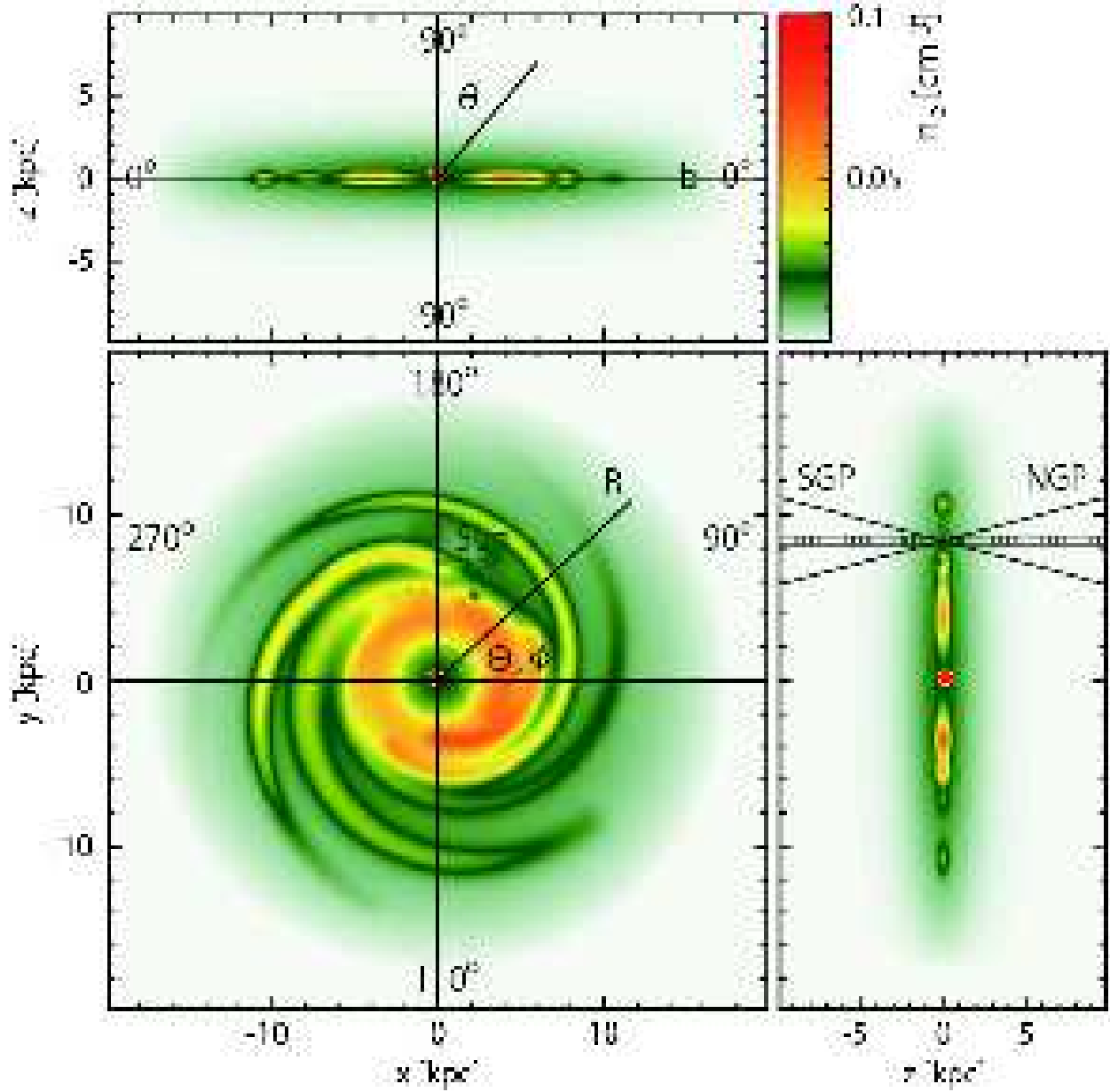}
\caption{
Electron density distribution, and definitions of the coordinate systems used in this paper. Small boxes in the bottom-right panel depicts the configuration of the data stacking (Section 2.2.2).
\label{figure1}
}
\end{figure}

\placefigure{figure1}

\begin{figure}[tp]
\figurenum{2}
\epsscale{1.0}
\plotone{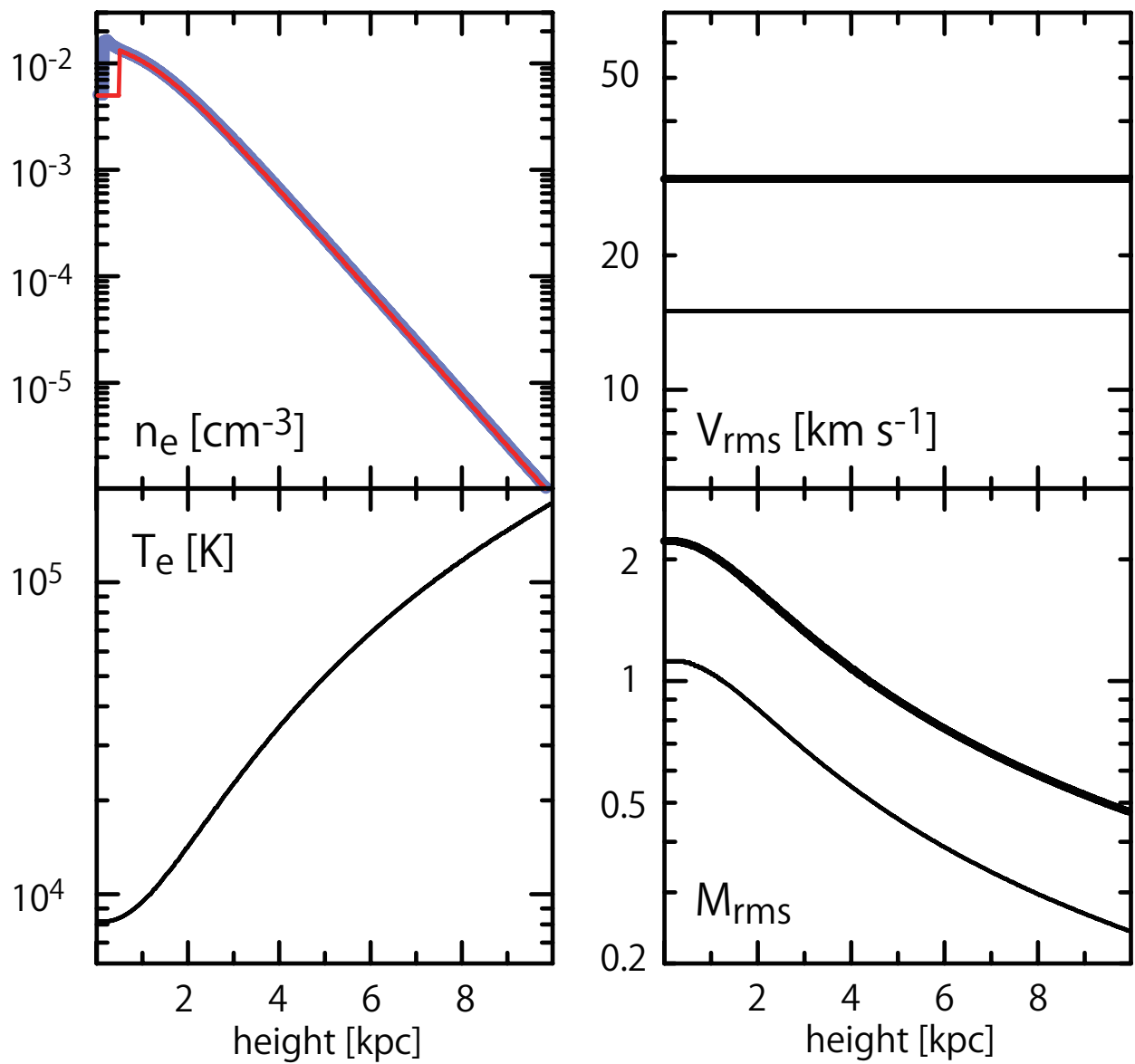}
\caption{
One-dimensional profiles from the Sun along the Galactic poles. Shown are the regular electron density (top left, thin red and thick blue are toward the NGP and SGP, respectively), the electron temperature (bottom left), the rms speed of random flow motions (top right), and the rms Mach number (bottom right, thin and thick lines are for $V_{\rm rms}=15$ and 30 ${\rm km~s^{-1}}$, respectively).
\label{figure2}
}
\end{figure}

\placefigure{figure2}

We first briefly describe our models for the global, regular components of the electron density and the GMF. The details along with the definitions of the coordinate systems used in the paper are described in Appendices. We also briefly describe models for the electron temperature and the rms speed of random flow motions, which are used in modeling the electron density fluctuations and the turbulent, random magnetic field in the next subsection.

For the electron density, $n_{e,0}$, we employ the NE2001 model \citep{cl02,cl03} displayed in Figure \ref{figure1}. We take the original parameters of the model except for the scale height of the thick disk, $h_1$, and the mid-plane electron density for the thick disk, $n_{e,1}$; we use $h_1=1.8~{\rm kpc}$ and $n_{e,1}=0.014~{\rm cm^{-3}}$ \citep{gmcm08}, which better reproduce both the dispersion measure (DM) and emission measure (EM) toward high Galactic latitude in our model (Appendices). The top-left panel of Figure \ref{figure2} shows one-dimensional profiles of $n_{e,0}$ from the Sun along the NGP and SGP. The electron density in the northern sky is smaller than that in the southern sky at low altitudes. This is due to the presence of a local, hot bubble with $n_{e,0} = 0.005~{\rm cm^{-3}}$, centered at $(x,y,z)=(0.01,8.45,0.17)$ in kpc \citep{cl02}.

\placefigure{figure3}

For the regular magnetic field, $\textit{\textbf{B}}_0$, we use combinations of the axi-symmetric spiral (ASS) or bi-symmetric spiral (BSS) field introduced by \citet{sun08}, the halo toroidal field of \citet{sr10}, and the halo poloidal fields of \citet{gkss10} or \citet{jf12}. The top panels of Figure \ref{figure3} show one-dimensional profiles of the strength of the regular field, $B_0$, and the line-of-sight (LOS) field strength, $B_{\parallel}$, from the Sun along the Galactic poles. For instance, ADPN indicates a model toward the NGP (N) including the ASS field (A) and the dipole toroidal field (D) with the dipole poloidal field (P). For the full list of models and the model name convention, refer to Table 1. We note that the dipole and quadrupole models are identical toward the NGP, but differ toward the SGP. The DS models tend to have larger $B_0$ than other models, since in other models the spiral and toroidal fields partly cancel each other. The spiral field dominates $B_0$ at $|z|<1.25~{\rm kpc}$, while the toroidal field dominates at $|z|\ge 1.25~{\rm kpc}$. For the LOS field strength, the field strength is $B_{\parallel} \simeq 0.2-0.3~\mu{\rm G}$ if the poloidal field exists. Otherwise, $B_{\parallel} = 0$.

For the electron temperature distribution, we employ the analytic expression adopted by \cite{sun08}
\begin{equation}
T_e(R,z)=5780+287R-526|z|+1770z^2,
\end{equation}
where $T_e$ is in units of K and the Galactocentric cylindrical coordinates $R$ and $z$ are in kpc. The bottom-left panel of Figure \ref{figure2} shows the one-dimensional profile from the Sun along the Galactic poles. 

For the rms speed of random flow motions, $V_{\rm rms}$, ${\rm H}\alpha$ observations provide a weak constraint on the plausible range of values, $V_{\rm rms}\sim 15-50~{\rm km/s}$ \citep{trh99,haf05,hil08,haf10}. \citet{hil08} studied the distribution of emission measure (EM), and found that the rms Mach number $M_{\rm rms} \equiv V_{\rm rms}/c_s \sim 1.4-2.4$ for $|b|>10^\circ$ and $M_{\rm rms}$ is smaller at higher Galactic latitudes. Studies of polarization gradients also broadly constrain $M_{\rm rms} \sim 0.5-2$ \citep{gae11,bur12}. We employ the simplest case, i.e., a uniform distribution of $V_{\rm rms}=15~{\rm km/s}$ ($M_{\rm rms}\sim 0.2-1$) or $V_{\rm rms}=30~{\rm km/s}$ ($M_{\rm rms}\sim 0.5-2$), as shown in the right panels of Figure \ref{figure2}. The bottom panels of Figure \ref{figure3} show one-dimensional profiles of $\beta_0$, the ratio of the gas pressure to the magnetic pressure due to $\textit{\textbf{B}}_0$. Our $\beta_0$ is in the range of $\sim 0.01-100$. Here, we assume that hydrogen is fully-ionized, helium is neutral, and that their mass fractions are $X=0.76$ and $Y=0.24$, respectively.

\subsection{Random Components}
\label{section2.2}

Observations suggest the presence of electron density fluctuations and turbulent magnetic fields, in addition to smooth components in the Galaxy. For instance, the volume filling factor of electrons, $\sim 0.05-0.5$, which quantifies the clumpiness, has been estimated from dispersion measures (DMs) and emissions/absorptions \citep[e.g.,][]{pw02, ber06, hil08, gmcm08, sun08}. Highly disturbed distributions of RM and polarization angle clearly indicate turbulent structures of the GMF \citep[e.g.,][]{sun08, tss09, wae09, gae11}. 

The random, turbulent components of the electron density and the GMF could be modeled analytically with preassigned spectra and random phases in Fourier space. As mentioned in Section 1, \citet{sr09} followed such an approach with a constant amplitude everywhere. \cite{jaf10} and \cite{jf12} introduced an ``ordered" or ``striated" field component. These works successfully reproduced the Galactic RM mainly toward low and mid-Galactic latitudes. However, it is not clear whether such treatments are good enough for studies of the Galactic RM toward high Galactic latitudes, where the random components are the dominant contribution to the RM. And constant rms amplitudes for electron density fluctuations and turbulent magnetic field would not be justified for broad distributions of $M_{\rm rms}$ and $\beta_0$ toward high altitudes (see Figures \ref{figure2} and \ref{figure3}), since they depend on $M_{\rm rms}$ and $\beta_0$ (see Appendix). In addition, in turbulent flows, phases are not really random.

An alternative approach would be to use the random electron density given by the NE2001 model for the electron density fluctuations. But we would then need to model the turbulent magnetic field separately. We therefore do not adopt this approach.

As in \citet{hil08}, we model the random components with MHD turbulence simulations in a closed box, as described below. In principle, if we performed full MHD simulations of the Galaxy, we could reproduce the electron density fluctuations and the turbulent magnetic field as well as the regular components in the Galactic disk and halo. But currently available computational resources do not allow a numerical resolution high enough to simultaneously reproduce both the large-scale global components and the small-scale turbulent components.

\subsubsection{MHD Turbulence Simulations}
\label{section2.2.1}

We embed the data of MHD turbulence simulations in the Galactic halo described by the global distributions of $n_{e,0}$ and $\textit{\textbf{B}}_0$ of the previous subsection. For this purpose, we carried out three-dimensional simulations of driven, isothermal, compressible MHD turbulence without self-gravity, using a multi-dimensional MHD code based on the Total Variation Diminishing (TVD) scheme \citep{krjh99}. This is an MHD extension of the explicit, second-order finite-difference, upwinded, conservative scheme of \citet{har83} for hydrodynamics. The version of the code used includes a flux constraining scheme that maintains $\nabla\cdot\textit{\textbf{B}}=0$ up to the machine accuracy \citep{ryu98}. In uniform, static medium with $\textit{\textbf{B}}_0$ assumed to be along the $x$-direction, turbulence was driven by imposing a solenoidal forcing with perturbations drawn from a Gaussian random field at wavenumbers around $k_{\rm drive}$ in Fourier space. The amplitude of the forcing was fixed in such a way that turbulence saturates at preassigned Mach numbers. Simulations were done in a periodic computational box with $512^3$ grid zones.

The size of the simulation box, $L_{\rm box}$, and the outer scale or the driving scale of turbulence, $L_{\rm drive} \equiv 2\pi/k_{\rm drive}$, were determined as follows. Observations suggest that $L_{\rm drive}$ is on the order of $\sim 1~{\rm pc}$ and $\sim 100~{\rm pc}$ in spiral arms and in interarm regions, respectively \citep{hbgm08}. These are ascribed to stellar sources such as stellar winds and protostellar outflows, or supernova and superbubble explosions. \citet{sr09} adopted $L_{\rm box}=10~{\rm pc}$ for studies of RM in low and mid-Galactic latitudes ($L_{\rm drive}$ is not defined in their approach). On the other hand, \citet{hil08} studied the EM distribution estimated with the Wisconsin ${\rm H}\alpha$ Mapper (WHAM). Using MHD turbulence simulations, they found that the EM distribution depends on $L_{\rm box}$ and $L_{\rm drive}$; larger $L_{\rm box}$ and $L_{\rm drive}$ results in smaller EMs. They argued that the histogram of the WHAM data on the warm ionized medium that includes high Galactic latitude data could be reproduced with $L_{\rm box}\sim 500$ pc and $L_{\rm drive}\sim 50-250$ pc. Adopting the results of \citet{hil08}, we set $L_{\rm box}=500~{\rm pc}$ and $L_{\rm drive} \simeq L_{\rm box}/2 = 250~{\rm pc}$. The grid of $512^3$ uniform zones for the box of $L_{\rm box}=500~{\rm pc}$ corresponds to the spatial resolution of $\sim 1~{\rm pc}$. We note that this resolution is enough to resolve the characteristic scales of turbulence; for instance, the most energy containing scale and the integral scale, $L_{kE(k)}\sim 50~{\rm pc}$ and $L_{\rm int}\sim 75~{\rm pc}$, respectively \citep[see][]{cr09}, are sufficiently large compared to the spatial resolution.

Representative simulations were performed for $M_{\rm rms}=0.5, 1, 2$ and $\beta_0=0.1, 1, 3, 10$, in order to cover the ranges of $M_{\rm rms}$ and $\beta_0$ toward the NGP and SGP (Figures \ref{figure2} and \ref{figure3}). Table 2 lists the simulations along with the rms values of density and magnetic field strength. For the case of $\beta_0 < 0.1$ at high Galactic altitudes ($|z|>5$~kpc, Figure \ref{figure3}), we use the data for $\beta_0=0.1$. Since the contribution from high altitudes to the integrated RM is at most several per cent, this approximation does not seriously affect our results. In turbulence simulations, $V_{\rm rms}$ saturates roughly in one flow-crossing time, $\sim L_{\rm box}/V_{\rm rms}$ \citep{wu12}. For the length scale of $500~{\rm pc}$, the corresponding timescale of $t\sim 1.6 \times 10^7~{\rm yr}$ and $\sim 0.8 \times 10^7~{\rm yr}$ for the rms flow speed of $15 ~{\rm km/s}$ and $30 ~{\rm km/s}$ is sufficiently short compared to the age of the Galaxy. We thus use the data at the saturation stage of turbulence.

\subsubsection{Construction of Model Space by Data Stacking}
\label{section2.2.2}

The data of simulations are stacked (or piled up) from the Galactic mid-plane up to the outer edge of the NE2001 model, $|z|=10.0~{\rm kpc}$ (Figure \ref{figure1}). In the stacking, we try to align the direction of the regular field in simulation data with that of the analytic model as follows. As we already noted, the regular field is dominated by the spiral field at $|z|<1.25~{\rm kpc}$ and by the toroidal field at $|z|\ge 1.25~{\rm kpc}$ (see Figure \ref{figure3}). In each domain, the regular field direction is close to either the $+x$ or $-x$ direction in our coordinate systems, except narrow transition regions with a width $\sim 0.2~{\rm kpc}$.

We divide each of the northern and southern hemispheres into four blocks:
\begin{eqnarray}
\left\{
\renewcommand{\arraystretch}{1.3}
\begin{array}{ll}
{\rm block~1}& (0.00\le |z|<1.25),\\
{\rm block~2}& (1.25\le |z|<2.50),\\
{\rm block~3}& (2.50\le |z|<5.00),\\
{\rm block~4}& (5.00\le |z|<10.0),\\
\end{array}
\renewcommand{\arraystretch}{1.0}
\right.
\end{eqnarray}
where $z$ is in units of kpc. We calculate the average values of $M_{\rm rms}$ and $\beta_0$ in each block and seek the simulation of closest parameters. Table 1 summarizes the direction of the regular magnetic field and the values of $M_{\rm rms}$ and $\beta_0$ adopted in each block. For example, in block 1 of ADON15, where 15 means $V_{\rm rms}=15~{\rm km/s}$, we choose the simulation with $M_{\rm rms}=1.0$ and $\beta_0=3.0$, and align the regular field with the $+x$-direction.

The stacked data are renormalized in such a way that we match the average electron density and magnetic field in the computational box with those of the regular components of Section 2.1. After the renormalization, each grid keeps the fluctuations based on MHD turbulence simulations.

\subsubsection{Goodness of the Modeling of Random Components}
\label{section2.2.3}

\placefigure{figure4}

Examples of profiles of the resulting electron density, magnetic field, and RM (see Section 2.3) from the Sun along the NGP, including both the regular and random components, are shown in Figure \ref{figure4}. It is apparent that fluctuations in the case with $V_{\rm rms}=30$ ${\rm km/s}$ are larger than those for $V_{\rm rms}=15$ ${\rm km/s}$, as expected. The strength of the turbulent magnetic field ($b = |\textit{\textbf{B}} - \textit{\textbf{B}}_0|$) is at most a few $\mu{\rm G}$, and mostly $\la 1~\mu{\rm G}$. This seems to be smaller than the strength of the random field, a few to several $\mu{\rm G}$, in the Galactic disk \citep[e.g., ][]{os93,beck96}, but consistent with $\sim 2~\mu{\rm G}$ a few kpc from the Galactic plane estimated by assuming equipartition between the thermal and nonthermal pressures \citep{cox05}, as well as with the recent estimates of $\le 1.5~\mu{\rm G}$ and $\le 1.4~\mu{\rm G}$ toward the NGP and SGP, respectively \citep{mao10}. The sign of LOS magnetic field changes several times. $|{\rm RM}|$ increases with the path length, and then saturates around $|z|\sim 2.0$~kpc where the electron density becomes small.

We check our model for the electron density distribution by comparing the resulting DM and EM with observed values \citep{pw02,hil08,gmcm08}. DM, which is determined only by the global, regular component, is 22 and 25 ${\rm pc~cm^{-3}}$ toward the NGP and SGP, respectively, while the observed value is $\sim 23-26$ ${\rm pc~cm^{-3}}$. We note that the NE2001 model, from which we get the regular component, was designed to reproduce observations including DM. The EM in our model is 1.5 (1.0) and 2.3 (1.6) ${\rm pc~cm^{-6}}$ toward the NGP and SGP, respectively, for $V_{\rm rms}=$ 30 (15) ${\rm km/s}$, which is in a good agreement with the observed value, $\sim 1-2$ ${\rm pc~cm^{-6}}$. Our model also reproduces the overall distribution of EM at $|b|>60^\circ$ presented by \citet{hil08}. Details are given in the Appendix.

\subsection{Rotation Measure Map}
\label{section2.3}

The RM toward a source outside the Galaxy is defined as
\begin{equation}\label{eq:RM}
{\rm RM}~({\rm rad~m^{-2}}) = 0.81 \int_{s_{\rm max}}^{0} n_{\rm e} B_\parallel ds,
\end{equation}
where $n_{\rm e}$, $B_\parallel$, and the path length $s$ are in units of ${\rm cm^{-3}}$, $\mu{\rm G}$, and pc, respectively. Conventionally RM is positive when the LOS magnetic field points toward us, and $s_{\rm max}$ is the maximum distance along the LOS up to the outer edge of our model space. In this paper, we calculate the RM only due to the GMF. We do not include other possible contributions, such as the RM due to the IGMF, the intrinsic RM at sources, and observational errors. We discuss those in Section 4.

We consider a square FOV of $900~{\rm deg^2}$ with $30^\circ$ on a side centered on each of the Galactic poles, which is comparable to the FOVs of RM observations toward the NGP and SGP \citep{mao10,sts11}. In addition, we also consider a square FOV of $200~{\rm deg^2}$ with $14.14^\circ$ on a side. The smaller FOV is roughly that proposed for the dense aperture array of the SKA \citep{fau10}. The smaller FOV is also used to check the FOV dependence of RM statistics.  To see structures in angular separations of $\gtrsim 0.3^\circ$ for which the second-order SF of observed RMs is available \citep{mao10,sts11}, we choose the angular resolution of pixels to be $\la 0.3^\circ$. The numbers of pixels we use are $N_{\rm pix}=256^2$ and $128^2$ for $900~{\rm deg^2}$ and $200~{\rm deg^2}$ FOVs, respectively.

We calculate RMs with Equation (\ref{eq:RM}) assuming one extragalactic radio source in each pixel, and construct mock RM maps. For each map, we randomly rotate stacked simulation boxes around the axis parallel to the regular magnetic field in Table 1 (keeping the direction of the regular magnetic field) and randomly shift box centers to avoid the repeat of the same grid zones which would make artifacts in statistics. When the integration reaches the top of a simulation box, the integration proceeds into the next stacked box and the integration is continued. If the integration reaches the side of the simulation box, we apply the periodic boundary condition to the side, and replicate the box beyond the side. 

We obtain 200 mock maps for each model listed in Table 1 and each FOV, and calculate statistical quantities. The quantities shown below in Sections 3.3 and 3.4 are the averages for 200 RM maps.

\section{Results}
\label{section3}

\subsection{Two-Dimensional Map}
\label{section3.1}

Figure \ref{figure5} shows sample maps over a $30^\circ \times 30^\circ$ FOV. Maps due to the regular components in the model ADON (leftmost panel) and due to the random components in one of 200 simulations for ADON30 (second panel) are shown. The combined map due to the regular and random components (third panel) as well as the binned map of median RMs in $2^\circ \times 2^\circ$ pixels (rightmost panel) are also shown. We can see that the regular components of density and magnetic field alone produce large-scale structures of up to the FOV size. The structures highlight $B_{\parallel}$ as induced by the radial and azimuthal components of the spiral and toroidal fields (the model shown does not include a poloidal field). The random components of density and magnetic field, on the other hand, produce complex structures, such as clump-like and filament-like features, on angular scales of $\sim 1 - 10^\circ$. Such structures persist in the combined map.

\citet{mao10} showed binned maps of median RMs in $2^\circ \times 2^\circ$ pixels toward the NGP and SGP, which were produced from observational data. The maps display structures of a few to several degrees. In our model, such structures are mostly the consequence of the random components of density and magnetic field  (rightmost panel). Our results indicate that the random components mainly produce the observed RM structures on scales less than several degrees, while the regular components contribute to larger scale structures.

\subsection{Contribution from Regular Components}
\label{section3.2}

We first examine the statistics of RMs in maps due to the regular components alone. We calculated the average, $\mu=\sum {\rm RM}/N_{\rm pix}$, and the standard deviation, $\sigma=\{\sum ({\rm RM}-\mu)^2/(N_{\rm pix}-1)\}^{1/2}$, in maps for the models and FOVs we consider. The resulting values are shown in Table 3. The average is mainly determined by the existence of the poloidal component of the GMF; $\mu \sim 0~{\rm rad~m^{-2}}$ without the poloidal component, or $\mu \sim -4.8~{\rm rad~m^{-2}}$ and $+5.6~{\rm rad~m^{-2}}$ toward the NGP and SGP, respectively, if the dipole poloidal component exists, and $\mu \sim -4.1~{\rm rad~m^{-2}}$ and $+4.7~{\rm rad~m^{-2}}$ toward the NGP and SGP, respectively, if the X-field poloidal component exists. The absolute values toward the NGP are a little smaller than those toward the SGP, due to the existence of a low density, local hot bubble centered in the northern sky (Figure \ref{figure2}). The averages for the $200~{\rm deg^2}$ FOV are slightly smaller than those for the $900~{\rm deg^2}$ FOV, since the smaller FOV includes high latitude regions only where the density and magnetic field strength are both smaller.

\placefigure{figure5}

The standard deviations are $\sigma \sim 0.2-0.9~{\rm rad~m^{-2}}$ and $\sim 0.7-4.3~{\rm rad~m^{-2}}$ toward the NGP and SGP, respectively, for the $900~{\rm deg^2}$ FOV. These are caused mostly by contributions from the radial and azimuthal components of the GMF (see also the leftmost panel of Figure \ref{figure5}). The DS models tend to have larger standard deviations than the DN and QS models, because the DS models have larger $B_0$ at mid and high altitudes as noted in Section 2.1. Larger values toward the SGP than those toward the NGP are again due to the existence of a local hot bubble near the Sun. The standard deviations for the $200~{\rm deg^2}$ FOV are smaller by a factor of $\sim 2$ than those for the $900~{\rm deg^2}$ FOV, since the smaller FOV includes a narrower range of RM values.

\subsection{Probability Distribution Function, Average, and Standard Deviation}
\label{section3.3}

\placefigure{figure6}

We next examine the statistics of RMs due to both the regular and random components. Figure \ref{figure6} shows the probability distribution functions (PDFs) of RMs in maps of $900~{\rm deg^2}$ FOV, averaged over 200 RM maps, for all models we consider. The PDFs roughly follow the Gaussian distribution, as pointed out by \citet{wu09}. The figure also shows the observed PDFs of \citet{mao10}. The average and standard deviation of $\mu$ (the average of RMs for a map) and $\sigma$ (the standard deviation of RMs for a map) over 200 maps were calculated. Figure \ref{figure7} shows the resulting values for $900~{\rm deg^2}$ FOV as well as for $200~{\rm deg^2}$ FOV.

\placefigure{figure7}

The value of $\mu$, which approximates the peak positions for the nearly-symmetric PDFs in Figure \ref{figure6}, is determined mostly by the regular components. So the average of $\mu$ is close to $\mu$ of the regular components alone in Section 3.1; $\langle \mu \rangle \sim 0$ ${\rm rad~m^{-2}}$, without the poloidal component of the GMF, or $\langle \mu \rangle \sim -5~{\rm rad~m^{-2}}$ and $\sim +6~{\rm rad~m^{-2}}$ toward the NGP and SGP, respectively, if the poloidal component exists. As noted in Section 1, the medians of observed RMs toward the NGP and SGP are $\sim 0.0\pm 0.5~{\rm rad~m^{-2}}$ and $\sim +6.3\pm 0.5~{\rm rad~m^{-2}}$, respectively \citep{mao10}. Hence, the models with poloidal components better reproduce the observed average toward the SGP, but the models without poloidal components are preferred for the NGP. On the other hand, the fluctuation of $\mu$ over the 200 maps is rather small; the standard deviations of $\mu$ for 200 maps are $\la 0.75~{\rm rad~m^{-2}}$ for models with $V_{\rm rms}=15~{\rm km~s^{-1}}$ and $\la 1.2~{\rm rad~m^{-2}}$ for models with $V_{\rm rms}=30~{\rm km~s^{-1}}$, as shown with error bars in Figure \ref{figure7}. These values are much smaller than the difference between the observed averages, $\sim 6.3$ ${\rm rad~m^{-2}}$, toward the NGP and SGP. This means that the difference cannot be explained by the statistical fluctuation caused by the random components in our models. These results indicate that none of our models can simultaneously reproduce the observed averages of RMs toward the NGP and SGP.

In Figure \ref{figure6}, the simulated PDFs have narrower and more sharply peaked profiles than the observed PDFs. The width of the PDFs is quantified by $\sigma$. In Figure \ref{figure7}, the averages of $\sigma$ are $0.9-1.3~{\rm rad~m^{-2}}$ toward the NGP and $1.3-4.3~{\rm rad~m^{-2}}$ toward the SGP for models with $V_{\rm rms}=15~{\rm km~s^{-1}}$, and $1.7-2.1~{\rm rad~m^{-2}}$ toward the NGP and $2.4-4.8~{\rm rad~m^{-2}}$ toward the SGP for models with $V_{\rm rms}=30~{\rm km~s^{-1}}$. The fluctuation of $\sigma$ in 200 maps is again small; the standard deviations of $\sigma$ for 200 maps are $\la 0.45~{\rm rad~m^{-2}}$ for models with $V_{\rm rms}=15~{\rm km~s^{-1}}$ and $\la 0.81~{\rm rad~m^{-2}}$ for models with $V_{\rm rms}=30~{\rm km~s^{-1}}$, as shown with error bars. On the other hand, the estimations of \citet{mao10} with their observed RMs are $\sigma \simeq 9.2~{\rm rad~m^{-2}}$ and $8.8~{\rm rad~m^{-2}}$ toward the NGP and SGP. So our estimations of $\sigma$ with simulated RMs are substantially smaller than the values of \citet{mao10}. We argue that all of our models fail to reproduce not only the peak positions but also the widths of the PDFs of observed RMs toward the NGP and SGP (see further discussions in Sections 4 and 5).

From Figure \ref{figure7}, we see the average and standard deviation of $\mu$ are not sensitive to the size of FOV. But the values of $\sigma$ for the $200~{\rm deg^2}$ FOV are somewhat smaller than those for the $900~{\rm deg^2}$ FOV, as expected. 

\subsection{Power Spectrum and Structure Function}
\label{section3.4}

\placefigure{figure8}

We also calculated the two-dimensional power spectrum (PS) and structure function (SF) with the sky map of RM. They tell us at which angular scales most power of RM resides and the spatial structure of RM decorrelates.

Figure \ref{figure8} shows the PS of RMs due to both the regular and random components in maps of $900~{\rm deg^2}$ FOV, averaged over 200 RM maps, for all models we consider. The PS toward the SGP are larger than those toward the NGP. This is again attributed to the presence of a local hot bubble in the northern sky. Toward the SGP, the PS for the models with a dipole toroidal field (D) are larger than those for the models with quadrupole toroidal field (Q). This is because the dipole models have larger $B_0$, as shown in Figure \ref{figure3}. (The dipole and quadrupole models are identical toward the NGP, as noted in Section 2.1.) The amplitude of the PS depends on the rms flow speed; the PS for $V_{\rm rms} = 30~{\rm km~s^{-1}}$ are larger by up to a factor of $\sim 3$ than those for $V_{\rm rms} = 15~{\rm km~s^{-1}}$. The slope also depends on the rms flow speed. Larger $V_{\rm rms}$'s result in shallower profiles. This is because the density and magnetic field PS of supersonic flows ($V_{\rm rms}=30~{\rm km/s}$ corresponds to $M_{\rm rms}\sim 1-2$) are shallower than those of subsonic flows ($V_{\rm rms}=15~{\rm km/s}$ corresponds to $M_{\rm rms}\sim 0.5-1$) \citep[e.g., see][]{pad04}. We quantified the slope of PS, $\alpha$, defined with $P(k) \propto k^{\alpha}$ in the angular range of $\sim 1 - 10^\circ$, and show the resulting values in Figure \ref{figure9} for $900~{\rm deg^2}$ FOV as well as for $200~{\rm deg^2}$ FOV. These are the averages for 200 maps. The Figure confirms the dependence of $\alpha$ on $V_{\rm rms}$; $\alpha$'s for $V_{\rm rms}=30~{\rm km/s}$ are larger by up to $\sim 0.4$ than those for $V_{\rm rms}=15~{\rm km/s}$ for the models shown in Figure \ref{figure8} (with $900~{\rm deg^2}$ FOV). Some of the PS have slopes consistent with the Kolmogorov slope $-5/3$, but others have steeper slopes, indicating that the GMF model influences the slope of the PS.

\placefigure{figure9}

From the observer's point of view, the SF of RMs is a statistical quantity that is easier to obtain than the PS. The $n$-th order SF is defined as 
\begin{equation}
S_n(r)=\langle|RM(\textit{\textbf{x}}+\textit{\textbf{r}})-RM(\textit{\textbf{x}})|^n\rangle_\textit{\textbf{x}}
\end{equation}
with $r=|\textit{\textbf{r}}|$, where the subscript indicates the averaging over the data domain of $\textit{\textbf{x}}$. Figure \ref{figure10} shows the second-order SFs $(n=2)$ in maps of $900~{\rm deg^2}$ FOV, averaged over 200 RM maps, for all models we consider. The SFs monotonically increase with angular separation, and reach up to $\sim 10~{\rm rad^2~m^{-4}}$ toward the NGP and $\sim$ several $\times 10~{\rm rad^2~m^{-4}}$ toward the SGP, respectively, at the angular separation of $\sim 10^\circ $. The SFs toward the SGP are larger than those toward the NGP. Toward the SGP, the SFs for the models with dipole toroidal fields are larger than those for the models with quadrupole toroidal fields. As for the PS, the SFs for $V_{\rm rms} = 30~{\rm km~s^{-1}}$ have amplitudes larger by up to a factor of $\sim 3$ than those for $V_{\rm rms} = 15~{\rm km~s^{-1}}$.

\placefigure{figure10}

Figure \ref{figure10} also shows the observed SFs toward the NGP and SGP \citep{sts11}. It is clear that the SFs of our simulated RMs do not match the observed SFs at any angular scales. At the largest observed scale of $\sim 10^\circ $, the simulated SFs are smaller by an order of amplitude or so than the observed SFs. But the difference is even larger at smaller angular separations. Not just the amplitudes, but also the slopes of the SFs of our simulated RMs are quite different from the observed ones; the slopes of simulated RMs are much steeper. We quantify the slope of SF, $\zeta$, defined with $S_2(r) \propto r^{\zeta}$ over the angular separations of $\sim 1 - 10^\circ$, and show the resulting values in Figure \ref{figure9} for a $900~{\rm deg^2}$ FOV as well as for a $200~{\rm deg^2}$ FOV. Again these are the averages for 200 maps. For the $900~{\rm deg^2}$ FOV, $\zeta \simeq 0.6 - 0.85$ for $V_{\rm rms} = 15~{\rm km~s^{-1}}$, and $\zeta \simeq 0.4 - 0.8$ for $V_{\rm rms} = 30~{\rm km~s^{-1}}$. The slopes of observed SFs are $\sim 0.02-0.05$ \citep{sts11}. So our models fail to reproduce the observed SFs not only in the amplitude but also in the slope. The results suggest that the observed RMs may contain contributions due to structures smaller than those typically found in the Galactic halo.

The PS and SFs for the $200~{\rm deg^2}$ FOV (not shown) show behaviors similar to those for the $900~{\rm deg^2}$ FOV, except that there are no data beyond a scale of $\sim 7^\circ $. Quantitatively, the slopes of PS and SF, $\alpha$ and $\zeta$, for the $200~{\rm deg^2}$ FOV are a bit smaller than those for the $900~{\rm deg^2}$ FOV, as shown in Figure \ref{figure9}.

\section{Discussion}
\label{section4}

In our models, the average or median of the RM toward high Galactic latitudes, $\mu$, is determined mostly by the halo poloidal component of the GMF, while the standard deviation, $\sigma$, is determined mostly by the random component of the GMF.

Our models fail to simultaneously reproduce the medians of observed RMs toward the NGP and SGP, $\sim 0.0\pm 0.5~{\rm rad~m^{-2}}$ and $\sim +6.3\pm 0.5~{\rm rad~m^{-2}}$ \citep{mao10}. The models that contain a poloidal component of the GMF have a vertical magnetic field of $|B_{\parallel}| \sim 0.3~\mu{\rm G}$ in the Earth vicinity, which induces $\langle \mu \rangle \simeq -5\pm (0.3-0.5)~{\rm rad~m^{-2}}$ toward the NGP and $\langle \mu \rangle \simeq 6\pm (0.6-1.1)~{\rm rad~m^{-2}}$ toward the SGP. The models without a poloidal component have $B_{\parallel} \simeq 0~\mu{\rm G}$, so $\langle \mu \rangle \simeq 0\pm (0.4-1.2)~{\rm rad~m^{-2}}$. The different values, $|B_{\parallel}| \simeq 0~\mu{\rm G}$ toward the NGP and $\sim 0.3~\mu{\rm G}$ toward the SGP, which would explain the observed medians, however, is not accommodated in our models. The difficulties of mixed regular field geometries in the steady state are discussed by \citet{mao10}. 

The fluctuation (standard deviation) of $\mu$ in simulations, $\la 1.2~{\rm rad~m^{-2}}$, is too small to account for the difference between the observed medians toward the NGP and SGP. To see whether larger fluctuations of $\mu$ are possible with different model parameters, simulations with larger values of $L_{\rm drive}$ and $V_{\rm rms}$ were carried out (not shown). We found that the fluctuation of $\mu$ increases with increasing $L_{\rm drive}$ and becomes $\sim 5~{\rm rad~m^{-2}}$ for $L_{\rm drive} \ga 1.5 - 2~{\rm kpc}$. However, with such large value of $L_{\rm drive}$, the EM distribution estimated from WHAM data \citep{hil08} cannot be explained (see Appendix). The fluctuation of $\mu$ increases with increasing $V_{\rm rms}$ also. We find a sufficiently large fluctuation for $V_{\rm rms} \ga 100~{\rm km~s^{-1}}$, which corresponds to $M_{\rm rms}\sim 6-8$. But such large $M_{\rm rms}$ is clearly inconsistent with the value constrained from the EM distribution \citep{hil08} and from polarization gradients \citep{gae11,bur12}.

Our models also fail to reproduce the standard deviations of observed RMs, $\sigma \simeq 9.2~{\rm rad~m^{-2}}$ and $8.8~{\rm rad~m^{-2}}$ toward the NGP and SGP, respectively \citep{mao10}. Our estimates, $\langle \sigma \rangle \la 4.5\pm(0.2-0.8)~{\rm rad~m^{-2}}$, are substantially smaller. The values are not consistent even with the Galactic contributions estimated by \citet{sch10}, $\sigma \sim 6.8\pm 0.1~{\rm rad~m^{-2}}$ and $8.4\pm 0.1~{\rm rad~m^{-2}}$ toward the NGP and SGP, respectively. (Note that the data of \citet{mao10} and \citet{sch10} are different.) We would get $\langle \sigma \rangle \sim 10~{\rm rad~m^{-2}}$ if we adopt $V_{\rm rms} \sim 100~{\rm km~s^{-1}}$. But again such a value is too large to explain the observed EM distribution.

We could explain the observed medians and standard deviations if there are additional global or local structures not represented in our models for the GMF \citep[e.g.,][]{mao10,sts11}. Such structures could have been produced, for instance, by the Parker instability or supernova explosions. In addition, \citet{mac12} have shown that transient structures can be produced in three-dimensional MHD simulations of turbulent gaseous disks. We could also address the discrepancy between the simulations and the data if the regular field is stronger than that in our models. For instance, we would have simulated the observed $\sigma$'s, if the regular field strength was increased by a factor of $\sim 2$. But then, the corresponding regular field strength near the Galactic plane would be $\sim 4~\mu{\rm G}$ (see Figure~\ref{figure3}), which is larger than that adopted in recent observational studies \citep{ptkn11,van11,jf12}. On the other hand, after investigating RMs toward the Perseus arm region, \citet{mao12} argued that observed RMs are consistent with a toroidal field of strength $\sim 2~\mu{\rm G}$ toward the north, but $\sim 7~\mu{\rm G}$ toward the south (note that the observation covers a rather thin region of $8.8~{\rm kpc}\le R \le 10.3~{\rm kpc}$ and $0.8~{\rm kpc} \le |z| \le 2.0~{\rm kpc}$). This indicates that there may be room for additional modeling of the GMF, which we leave for future studies.

Even with additional structures and stronger regular fields, however, the observed SFs are difficult to reproduce \citep{sts11}. The second-order SFs of simulated RMs are substantially smaller than the observed SFs, especially at small angular separations of $0^\circ.1 - 1^\circ$. This means that we need structures that would provide significant powers on scales of a parsec or so, but there is no observational support for such structures \citep[e.g.,][]{gae05,hbgm08}. The slope of observed SFs, $0.02-0.05$ \citep{sts11}, is much flatter than that of simulated SFs, $\sim 0.4-0.9$. We note that two-dimensional, white noise results in a flat SF; $S_2\sim 2\sigma_{\rm err}^2$, where $\sigma_{\rm err}$ is the standard deviation of the noise. We would get $S_2\sim 200~{\rm rad^2~m^{-4}}$ for $\sigma_{\rm err}\sim 10~{\rm rad~m^{-2}}$. Observational errors should follow the distribution of white noise. But \citet{mao10} estimated that the observational errors in their data are $\sim 5$ and $\sim 3~{\rm rad~m^{-2}}$ toward the NGP and SGP, respectively. \citet{tss09} estimated that the error in their data is $\sim 8~{\rm rad~m^{-2}}$. So observational errors may not be enough to explain the observed flat SFs of $\sim 100-300~{\rm rad^2~m^{-4}}$ \citep[see also][]{ham12}.

It is expected that radio sources against which RMs are observed have their own intrinsic RMs. The RMs of radio sources should also follow the distribution of white noise and contribute to a flat SF. \citet{scs84} and \citet{sc86} showed that the observed SF toward the NGP deviates from a flat behavior at very small scales; the square value of RM difference between two sources at $\textit{\textbf{x}}_i$ and $\textit{\textbf{x}}_j$, $\{RM(\textit{\textbf{x}}_i)-RM(\textit{\textbf{x}}_j)\}^2$ (not the second-order SF), is around $\sim 10~{\rm rad^2~m^{-4}}$ at angular separations of $\sim 0.01^\circ$. This tells us that even if the square value is solely due to the RMs of radio sources, it is unlikely that the RMs make up the observed flat SFs of $\sim 100-300~{\rm rad^2~m^{-4}}$. In fact, \citet{sc86} claimed that many observed sources possess little intrinsic RM from the fact that the fractional polarization percentage is nearly constant with increasing wavelength.

Finally, there exists clear evidences for an extragalactic component to the RMs, from three different papers taking different approaches \citep{sch10,ber12,ham12}. Although \cite{ber12} and \cite{ham12} attribute the excess scatter in RM to individual intervening absorbers along the line of sight, an alternative origin which can contribute to the observed RM toward high Galactic latitudes too is the IGMF. As mentioned in Section 1, using simulations for the large-scale structure formation in the universe, \citet{ar11} predicted that the RM due to the IGMF would have $\mu \simeq 0$ and $\sigma \simeq$ several ${\rm rad~m^{-2}}$. They also predicted that the SF of the RM has a flat profile of $100-200~{\rm rad^2~m^{-4}}$ amplitude at angular separations larger than $0.2^\circ$; the SF is expected to decrease to the order of $\sim 10~{\rm rad^2~m^{-4}}$ at $\sim 0.01^\circ$ from $\sim 100~{\rm rad^2~m^{-4}}$ at $\sim 0.1^\circ$, consistent with the observation of \citet{sc86}. The work of \citet{ar11} suggests the possibility that a substantial fraction of the RM toward the NGP and SGP can be attributed to the RM due to the IGMF. If so, the observed standard deviations and SFs can be explained. But reproducing the observed medians of RM still needs additional components or/and structures of the GMF.

\section{Conclusion}
\label{section5}

We have studied the Galactic RM toward high Galactic latitudes. We have considered a number of models for the global, regular components of the GMF and the electron density in the Galaxy, based on observations. The turbulent, random components were modeled with three-dimensional MHD turbulence simulations. The strength of the regular magnetic field in our models is a few $\mu$G close to the disk and smaller at high altitudes. The strength of the turbulent field is at most a few $\mu{\rm G}$, and mostly $\la 1~\mu{\rm G}$. We obtained RM maps for $900~{\rm deg^2}$ FOV toward the Galactic poles, and compared the results with observations. We also considered a smaller FOV of $200~{\rm deg^2}$, designed to simulate the FOV of future surveys with the SKA.

Our models fail to simultaneously reproduce the observed medians of RMs toward the NGP and SGP. The observations require vertical magnetic fields of $B_{\parallel} \sim 0~\mu{\rm G}$ toward the NGP and $0.3~\mu{\rm G}$ toward the SGP in the Earth's vicinity, but such field geometries are not accommodated in the GMF models we considered. The PDFs of simulated RMs are narrower and more sharply peaked than the observed PDFs, meaning that the standard deviations of simulated RMs are smaller than the observed values. The second-order SFs of simulated RMs are one to two orders of magnitude smaller than the observed SFs at small angular separations. In addition, the slopes of the SFs of simulated RMs are substantially larger than the observed ones.

We argue that observational errors and the intrinsic RM of background radio sources are not enough to explain the discrepancies between the statistics of our simulated RMs and observed RMs. We suggest that the RM due to the IGMF may account for a fraction of the RM toward high Galactic latitudes and could explain the discrepancies in the standard deviation and SF of the RMs. As a subsequent project, the RM due to the IGMF as well as observational errors and the intrinsic RM of background radio sources would be incorporated in the modeling of the RM toward high Galactic latitudes, to check quantitatively whether some of the discrepancies, such as those in the standard deviation and SF, would be explained. But we expect that reproducing the observed medians will still need additional components or/and structures of the GMF.

We should note, however, that current observations of RMs still contain large uncertainties. New and the future observational facilities such as the JVLA, MWA, LOFAR, ASKAP, MeerKAT, and the SKA will produce much better data. For instance, ASKAP will detect radio sources with average angular separations of $\sim 0.1^\circ$. With such a dense RM grid, the quality of the observed RM data will be dramatically improved. Better quality data will hopefully enable us to better quantify the contributions due to the GMF, the IGMF, the intrinsic RM of background radio sources, and observational errors.

Finally, we note that better measurements of the Mach number of turbulence in the halo by observations of ${\rm H_\alpha}$ line profiles \citep{haf05,hil08,haf10} as well as by radio polarization gradients \citep{gae11} will help us improve the constraints on the magnitude and structures of the RM toward high Galactic latitudes. 

\appendix

\section{Coordinate Systems}

The following coordinate systems are used throughout the paper: Cartesian coordinates, $(x,y,z)$, Galactocentric cylindrical coordinates, $(R,\Theta,z)$, Galactocentric polar coordinates, $(r,\theta,\phi)$, and Galactic celestial coordinates, $(l,b)$, respectively, defined in Figure \ref{figure1}. Here, $R=(x^2+y^2)^{1/2}$ is the Galactocentric radius, $\Theta=\phi$ is the azimuth angle starting from $l=90^\circ$ and increasing in the counterclockwise direction, and the $x$-$y$ plane coincides with the Galactic plane with $x$ pointing to $l=90^\circ$ and $y$ to $l=180^\circ$. The Sun is located at $(x,y,z)=(0,8.5,0)$ in kpc, and thus $R_\odot = 8.5$~kpc.

\section{Electron Density}

The NE2001 model we employ for the electron density, $n_{e,0}$, consists of three global components (thick disk, thin disk, and five spiral arms), four local components (local hot bubble, Loop I, local super bubble, and low density region in quadrant 1), and more than a hundred isolated components such as clumps and voids, as displayed in Figure \ref{figure1}. The electron density of the model was obtained from the average of the DM for pulsars at known distances. We take the original parameters in the NE2001 package\footnote{We use NE2001\_1.0 downloaded from http://astrosun2.astro.cornell.edu/{$\sim$}cordes/NE2001/.}, except for the scale height of the thick disk, $h_1$, and the mid-plane electron density for the thick disk, $n_{e,1}$. \citet{gmcm08} obtained $h_1 \simeq 1.8$~kpc from an analysis of pulsars at high Galactic latitudes. We adopt $h_1=1.8$~kpc, instead of $0.97$~kpc in the package. The corresponding mid-plane electron density for the thick disk is $n_{e,1}=0.014~{\rm cm^{-3}}$.

The revised scale height improves fits of the Galactic RM and radio continuum emission at low and mid-Galactic latitudes \citep{sr10}. The changes of the scale height and mid-plane electron density of the thick disk, however, may not be consistent with the construction of the NE2001 model, because these parameters are highly covariant with others in the model. In order to justify our modification and make sure the consistency with observations, we compare the DM and EM from the NE2001 model with the original scale height and mid-plane density (hereafter the original NE2001 model) and the NE2001 model with the modified scale height and mid-plane density (our modified NE2001 model). We also tested the exponential model fitted by \citet{gmcm08} (the plane-parallel model) for comparison.

Figure \ref{figure11} shows the distribution of DM as a function of height above the Galactic plane. At a height of $0.2-0.4$~kpc, both the original and modified NE2001 models reproduce the distribution of \citet{gmcm08} well toward the SGP, but predict smaller DM by a factor of $\sim 2$ toward the NGP, likely due to the local hot bubble. At the height of $ \ga 1$~kpc, the original NE2001 model overestimates DM by a factor of $\sim 1.5$ toward south. On the other hand, the modified NE2001 model reproduces the observed DM well; the resultant DMs for the modified NE2001 are $22~{\rm pc~cm^{-3}}$ and $25~{\rm pc~cm^{-3}}$ toward the NGP and SGP, respectively, which are roughly consistent with observations \citep{pw02,hil08,gmcm08}.

\placefigure{figure11}

Next, we calculated the PDF of EM. There is a well-known issue that models with smooth electron density profile do not reproduce the observed EM well; the discrepancy can be resolved by introducing volume filling factor, $f_{\rm e}$. We adopted $f_{\rm e}=0.07 \exp (|z|/0.5)$, $z$ in kpc, for $z \le 0.75$ kpc and $f_{\rm e}=0.32$ for $z > 0.75$ kpc \citep{ber06,sun08}. The resulting EM is shown in Figure \ref{figure12}. With the adopted $f_{\rm e}$, the original NE2001 does not correctly reproduce the EM distribution from the WHAM observation \citep{hil08}. On the other hand, the modified NE2001 model and the plane-parallel model produce results, which are better consistent with the observed distribution. 

\placefigure{figure12}

It is interesting to see that the cases with $V_{\rm rms}=15$~${\rm km~s^{-1}}$ (or the rms Mach number of turbulence is close to $\sim 1$) better reproduce the observed EM. It should be pointed that \citet{gmcm08} claimed smaller $f_{\rm e}$ (that is, more clumpy) at both low and high altitudes. If smaller $f_{\rm e}$ is adopted, our model gives larger dispersion of EM, so that even smaller $V_{\rm rms}$ would be preferable. Then, the corresponding standard deviation of RM would be even smaller (Figure \ref{figure7}).

Finally, Figure \ref{figure13} shows the second-order SF of simulated RMs in a 900 ${\rm deg^2}$ FOV. The amplitude of SF for the original NE2001 model is somewhat larger than that for the modified NE2001 and plane-parallel models. The slope of SF does not significantly depend on the electron density model, and again the SF for the three models does not match the observed one (see Section 3.4).

\placefigure{figure13}

\section{Magnetic Field}

The global, regular magnetic field, $\textit{\textbf{B}}_0$, is conventionally described as the combination of the disk spiral field, $\textit{\textbf{B}} _{\rm s}$, the halo toroidal field, $\textit{\textbf{B}}_{\rm t}$, and the halo poloidal field, $\textit{\textbf{B}}_{\rm p}$, so that $\textit{\textbf{B}}_0= \textit{\textbf{B}}_{\rm s}+\textit{\textbf{B}}_{\rm t}+\textit{\textbf{B}}_{\rm p}$ \citep[e.g., ][]{ps03}. There are a number of analytic models intended to reproduce the observed structure of $\textit{\textbf{B}}_0$. The models have been tuned by fitting the data mostly at low and mid-Galactic latitudes, but can be used for studies of the Galactic RM toward high Galactic latitudes as well. Specifically, we employ the models introduced by \citet{sun08} for the spiral field, by \citet{sr10} for the toroidal field, by \citet{gkss10} for the dipole poloidal field, and by \citet{jf12} for the X-field poloidal field, unless otherwise specified. We note that our combined model does not take account of detailed theoretical consistencies, such as the continuity, divergence-free nature, and closeness of $\textit{\textbf{B}}_0$. We expect that those do not significantly affect our results.

The large-scale disk fields of spiral galaxies including our Galaxy have been classified into two types based on the spiral pattern, the axi-symmetric spiral (ASS) field with no dependence on the azimuthal angle, or the bi-symmetric spiral (BSS) field with a symmetry of $\pi$. The functional form for the disk spiral field can be written as
\begin{eqnarray}
\left\{
\renewcommand{\arraystretch}{1.5}
\begin{array}{ll}
B_{{\rm s},R}(R,\Theta,z) & = D_1(R,z)D_2(R,\Theta)\sin (p_{\rm s0}),\\
B_{{\rm s},\Theta}(R,\Theta,z) & = -D_1(R,z)D_2(R,\Theta)\cos (p_{\rm s0}),\\
B_{{\rm s},z}(R,\Theta,z) & = 0,
\end{array}
\renewcommand{\arraystretch}{1.0}
\right.
\end{eqnarray}
where $p_{\rm s0}$ is the pitch angle of arms which is positive for leading spirals and negative for trailing spirals such as in our Galaxy, and
\begin{eqnarray}
D_1(R,z)=\left\{
\renewcommand{\arraystretch}{2.5}
\begin{array}{ll}
\displaystyle  B_{\rm s0}\exp{
\left( -\frac{R-R_\odot}{R_{\rm s0}}-\frac{|z|}{z_{\rm s0}}\right)
} & R>R_{\rm sc},\\
\displaystyle  B_{\rm sc}\exp{
\left(-\frac{|z|}{z_{\rm s0}}\right)
} & R \le R_{\rm sc}.
\end{array}
\renewcommand{\arraystretch}{1.0}
\right.
\end{eqnarray}

For the ASS model, we adopt the ASS + reversals of the magnetic field directions:
\begin{eqnarray}
D_2(R,\Theta)=\left\{
\renewcommand{\arraystretch}{1.3}
\begin{array}{ll}
+1 & ~~~R>7.5~{\rm kpc},\\
-1 & ~~~6~{\rm kpc}<R\le 7.5~{\rm kpc},\\
+1 & ~~~5~{\rm kpc}<R\le 6~{\rm kpc},\\
-1 & ~~~R\le 5~{\rm kpc},
\end{array}
\renewcommand{\arraystretch}{1.0}
\right.
\end{eqnarray}
where $+1$ means the clockwise direction as seen from the north pole. We also adopt $R_{\rm s0}=10~{\rm kpc}$, $z_{\rm s0}=1~{\rm kpc}$, $R_{\rm sc}=5~{\rm kpc}$, $B_{\rm s0}=2~{\rm \mu G}$, $B_{\rm sc}=2~{\rm \mu G}$, and $p_{\rm s0}=-12^\circ$.

For the BSS model, we adopt the spiral structure of
\begin{equation}
D_2(R,\Theta)=\sin \left(
\Theta+\frac{1}{\tan p_{\rm s0}}\ln\frac{R}{R_{\rm sb}}\right).
\end{equation}
We also adopt $R_{\rm s0}=6~{\rm kpc}$, $z_{\rm s0}=1~{\rm kpc}$, $R_{\rm sc}=3~{\rm kpc}$, $B_{\rm s0}=2~{\rm \mu G}$, and $B_{\rm sc}=2~{\rm \mu G}$. In addition, $R_{\rm sb}=9~{\rm kpc}$ and $p_{\rm s0}=-10^\circ$ for $R>6~{\rm kpc}$, and otherwise $R_{\rm sb}=6~{\rm kpc}$ and $p_{\rm s0}=-15^\circ$. Note that the trigonometric function is minus cosine if the azimuth angle, $\Theta$, starting from $l=180^\circ$ is adopted \citep[e.g.,][]{sun08}.

\begin{figure}[tp]
\figurenum{3}
\epsscale{1.0}
\plotone{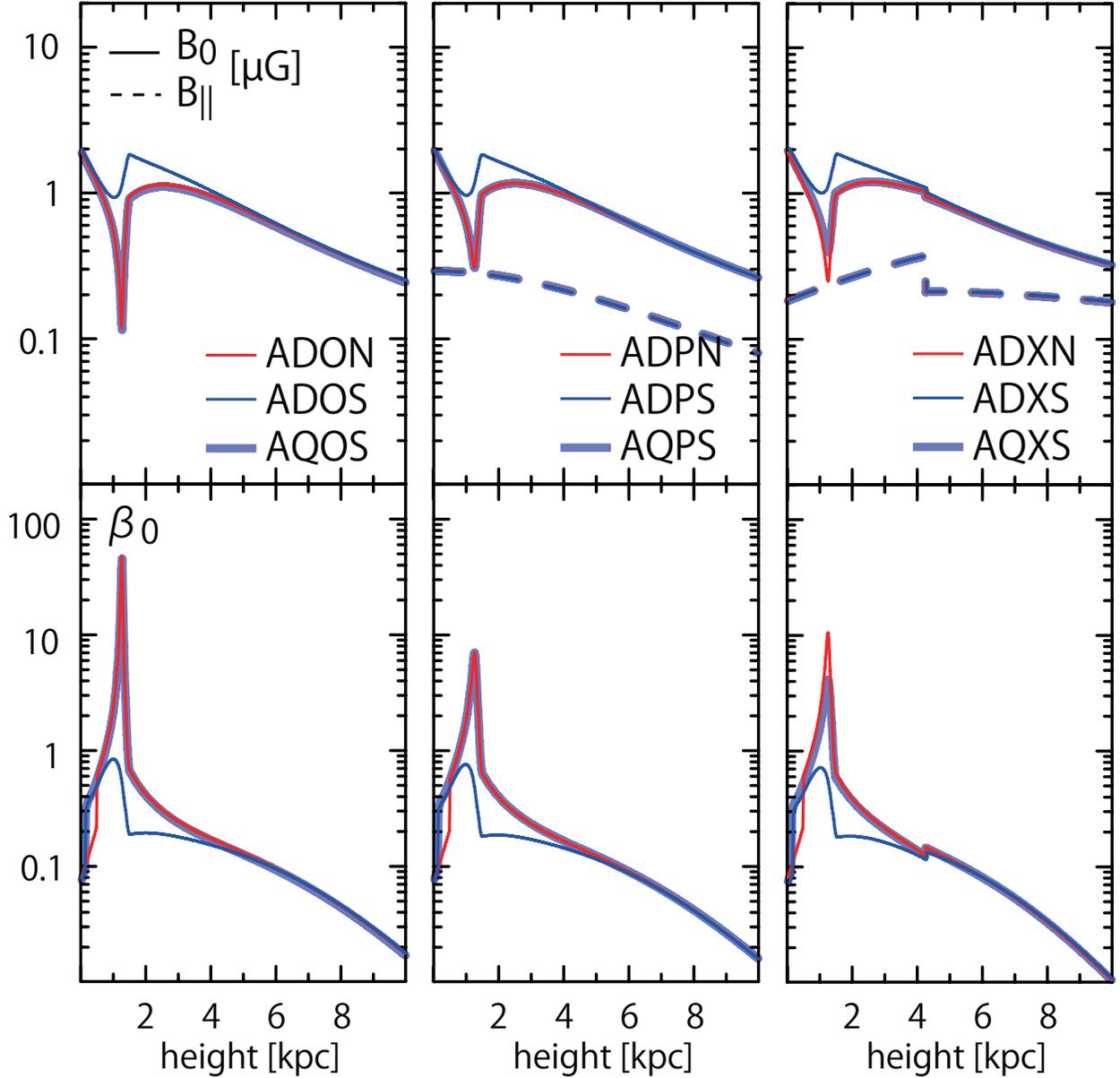}
\caption{
One-dimensional profiles from the Sun toward the Galactic poles. The top panels show the regular field strength, $B_0$ (solid), and the LOS field strength, $B_{\parallel}$ (dashed). $B_{\parallel}$ is pointing away from us and toward us for models toward the NGP and SGP, respectively. The bottom panels show the plasma beta of the regular magnetic field, $\beta_0$. Profiles only for axi-symmetric spiral models are shown. The names of models mean: (A) axi-symmetric spiral, (B) bi-symmetric spiral, (D) dipole toroidal, (Q) quadrupole toroidal, (O) no poloidal, (P) dipole poloidal, (X) X-field poloidal, (N) toward the NGP, and (S) toward the SGP.
\label{figure3}
}
\end{figure}

\begin{figure}[tp]
\figurenum{4}
\epsscale{0.7}
\plotone{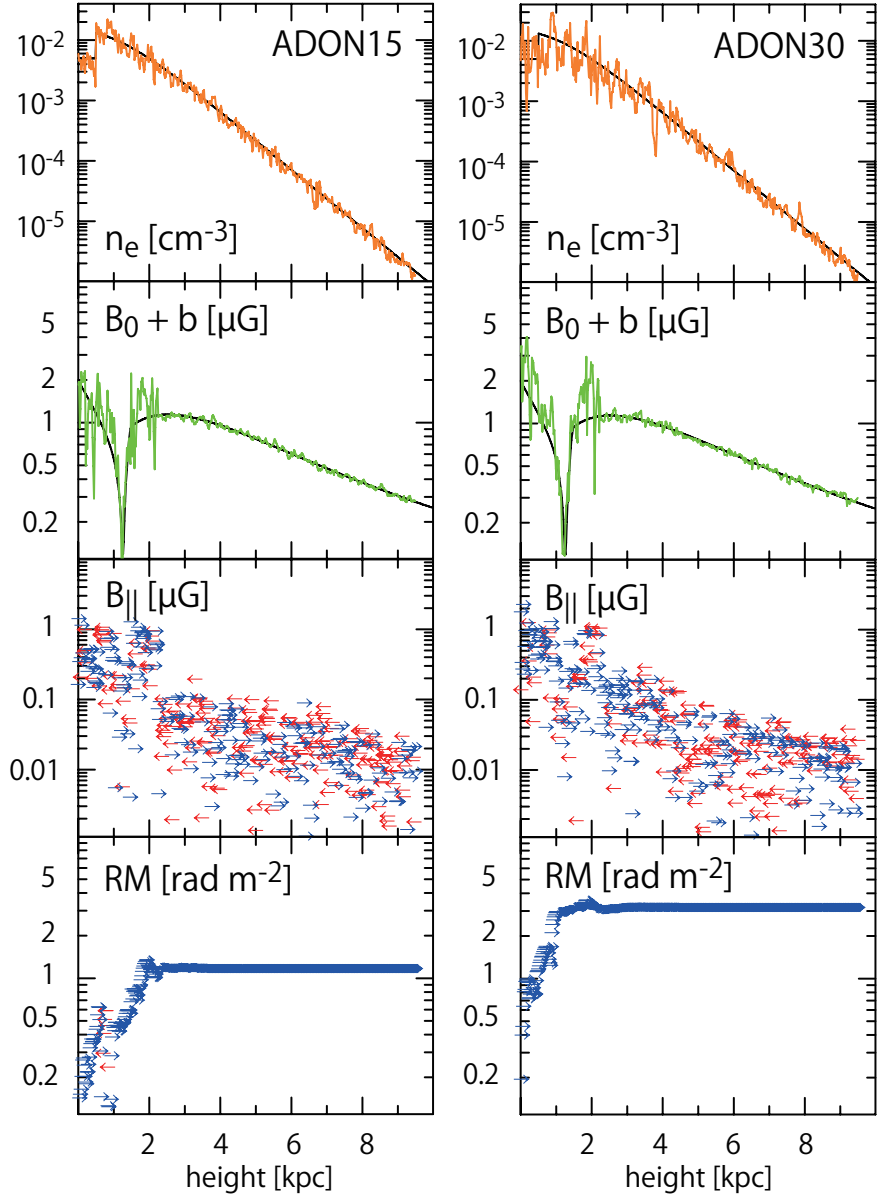}
\caption{
One-dimensional profiles from the Sun toward the north Galactic pole. The left and right panels show the profiles for ADON15 and ADON30, respectively. The Panels from top to bottom show the electron density, the total magnetic field strength, the LOS magnetic field strength, and the cumulative RM from the Sun, respectively. The arrows with red and blue colors indicate the direction of magnetic field and the sign of RM. In the top two panels, the black lines show the regular components. $B_{\parallel}$ of the regular magnetic field is zero along the Galactic poles in ADON15 and ADON30 without the poloidal component.
\label{figure4}
}
\end{figure}

\begin{figure*}[tp]
\figurenum{5}
\epsscale{1.0}
\plotone{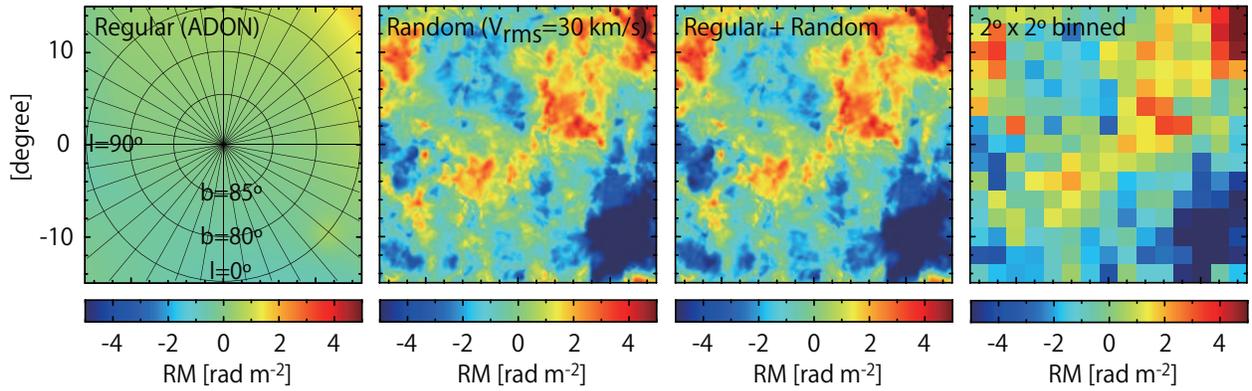}
\caption{
Two-dimensional RM map of the NGP with $30^\circ\times 30^\circ$ FOV for ADON30. Panels from left to right show RM maps due to: the regular components of the electron density and the GMF only; the random components only; both the regular and random components; and a map binned in $2^\circ$ by $2^\circ$ pixels. The Galactic celestial coordinates are shown in the leftmost panel.
\label{figure5}
}
\end{figure*}

\begin{figure*}[htbp]
\figurenum{6}
\epsscale{0.9}
\plotone{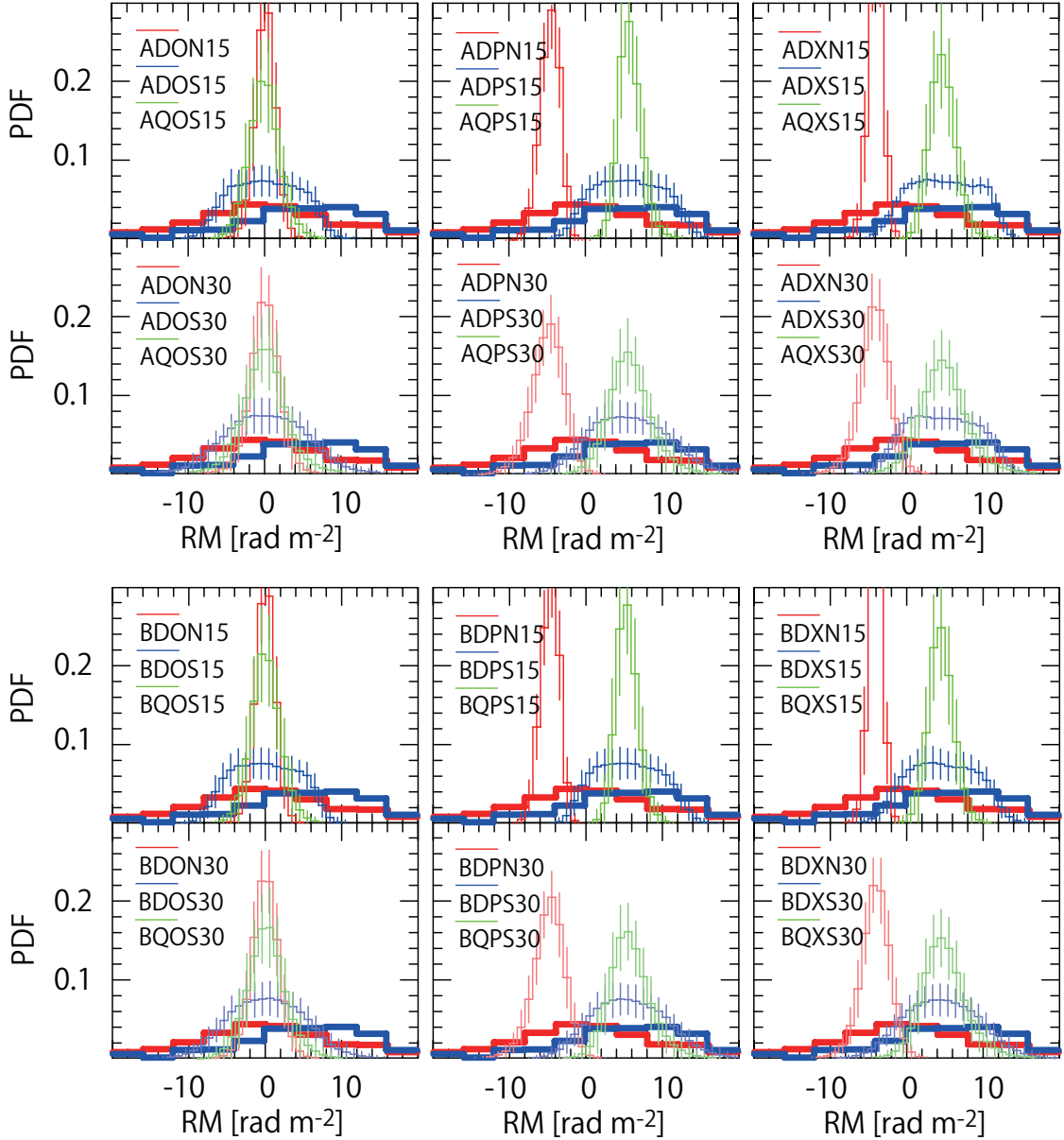}
\caption{
Probability distribution functions (PDFs) of simulated RMs in $30^\circ\times 30^\circ$ FOV for models we consider (thin lines). Shown PDFs are the averages for 200 maps, where error bars indicate the standard deviation. Also shown as thick lines are the PDFs of observed RMs toward the NGP (red) and SGP (blue), respectively \citep{mao10}. \label{figure6}
}
\end{figure*}

\begin{figure*}[tp]
\figurenum{7}
\epsscale{1.0}
\plotone{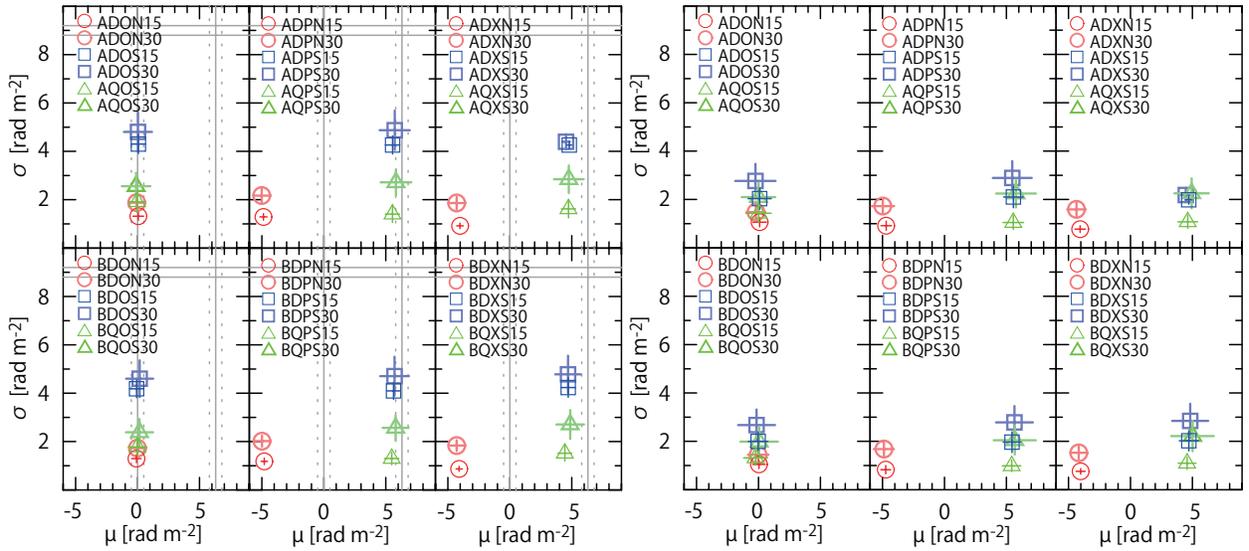}
\caption{
{Average $\mu$ and standard deviation $\sigma$ of simulated RMs in $30^\circ\times 30^\circ$ (left) and $14.14^\circ\times 14.14^\circ$ (right) FOVs for models we consider. Symbols are the averages of $\mu$ and $\sigma$ for 200 maps, while error bars indicate the standard deviations over 200 maps. Also shown as gray solid and dashed lines are observed values and errors: $\mu \sim 0.0\pm 0.5~{\rm rad~m^{-2}}$ and $\sigma \simeq 9.2~{\rm rad~m^{-2}}$ toward the NGP and $\mu \sim +6.3\pm 0.5~{\rm rad~m^{-2}}$ and $\sigma \simeq 8.8~{\rm rad~m^{-2}}$ toward the SGP \citep{mao10}.}
\label{figure7}
}
\end{figure*}

\begin{figure*}[tp]
\figurenum{8}
\epsscale{0.9}
\plotone{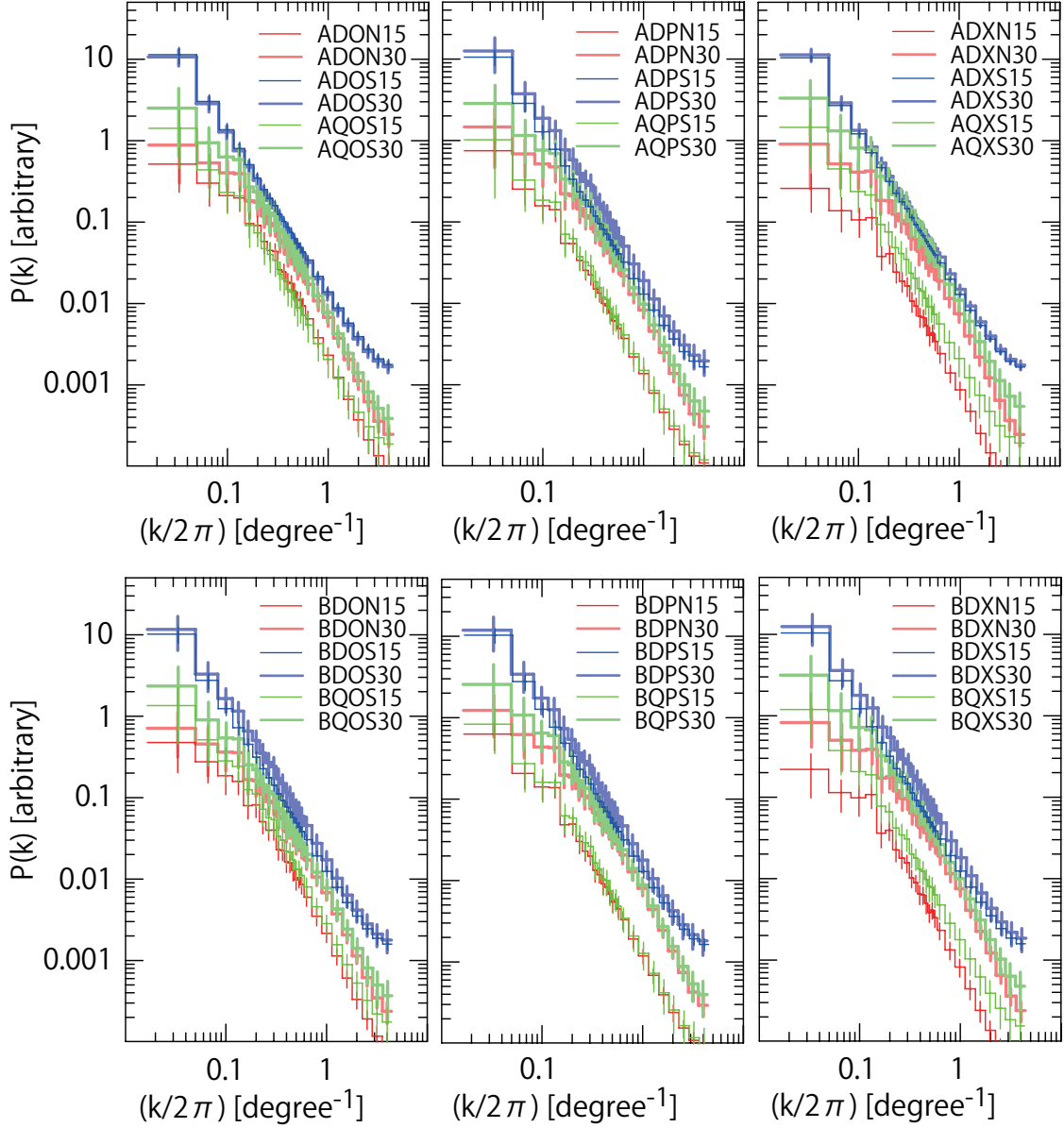}
\caption{
Power spectra (PS) of simulated RMs in $30^\circ\times 30^\circ$ FOV for models we consider. PS shown are the average over 200 maps, where error bars indicate the standard deviation. \label{figure8}
}
\end{figure*}

\begin{figure*}[tp]
\figurenum{9}
\epsscale{1.0}
\plotone{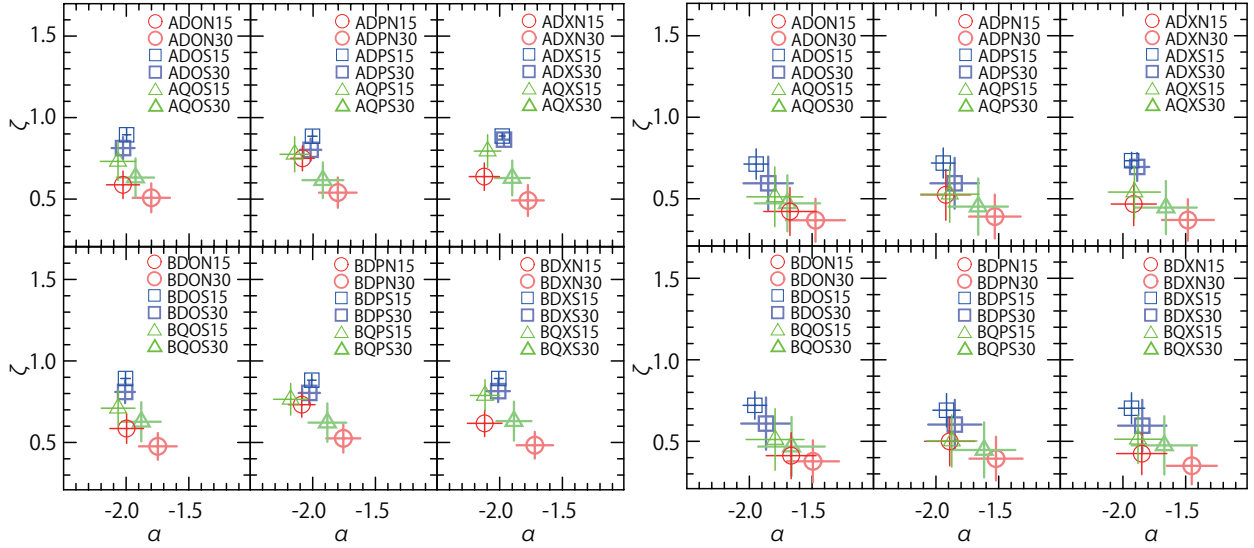}
\caption{
{ 
Slopes of PS ($\alpha$) and SF ($\zeta$) of simulated RMs over $30^\circ\times 30^\circ$ (left) and $14.14^\circ\times 14.14^\circ$ (right) FOVs for models we consider. Symbols are the averages over 200 maps, while error bars indicate the corresponding standard deviations. }
\label{figure9}
}
\end{figure*}

\begin{figure*}[tp]
\figurenum{10}
\epsscale{0.9}
\plotone{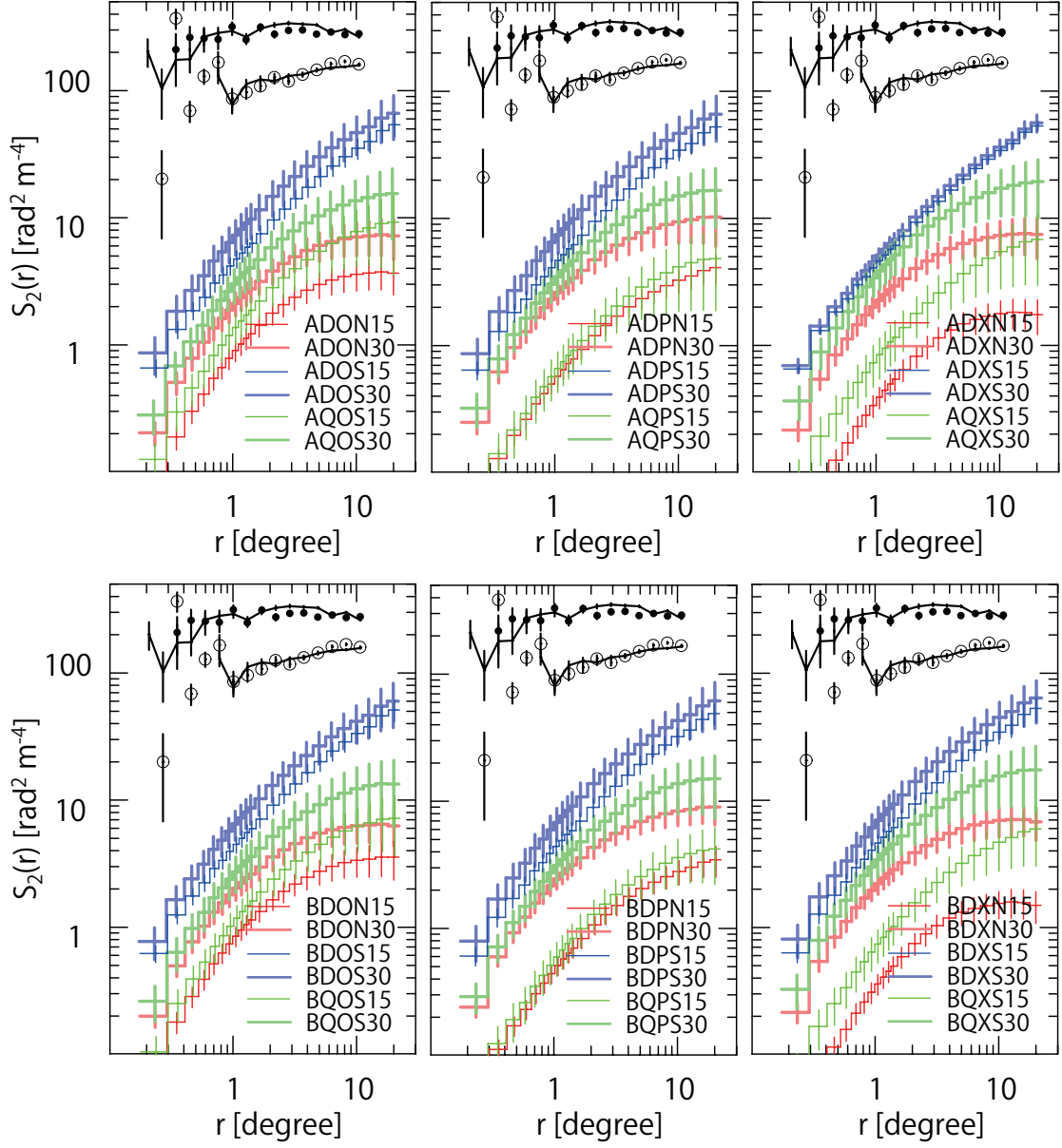}
\caption{
Second-order structure functions (SFs) of simulated RMs over a $30^\circ\times 30^\circ$ FOV for models we consider. Shown SFs are the averages for 200 maps, where error bars indicate the standard deviation for 200 maps. Also shown are the observed second-order SFs (\citet[][circles]{mao10} and \citet[][lines]{sts11}). Open circles and thick lines are toward the NGP, and filled circles and thin lines are toward the SGP. \label{figure10}
}
\end{figure*}

\begin{figure*}[tp]
\figurenum{11}
\epsscale{0.9}
\plotone{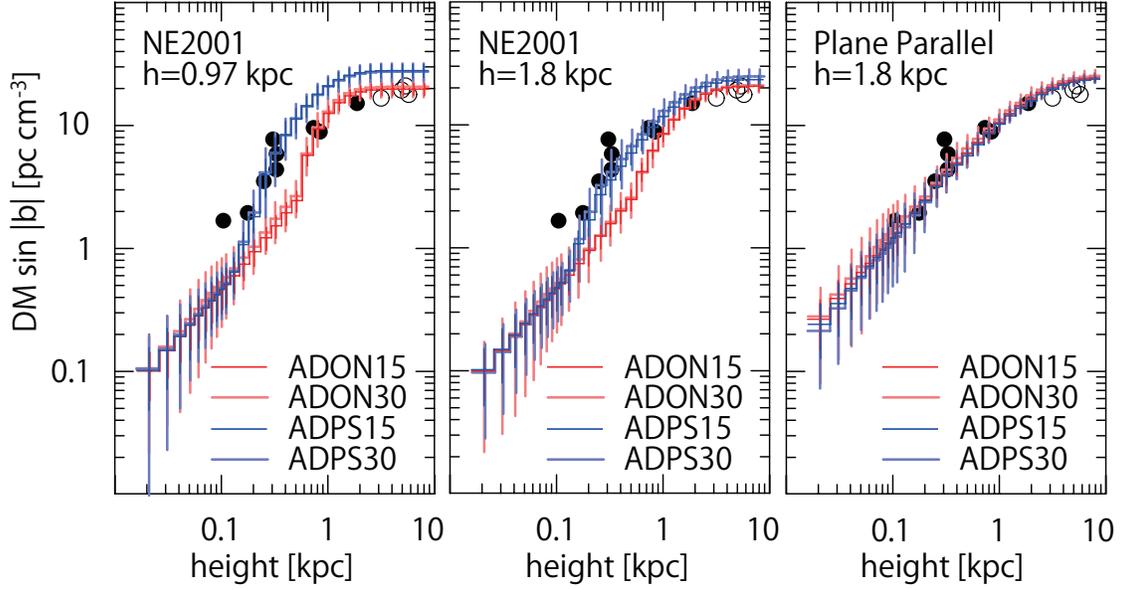}
\caption{
{Distributions of DM as a function of height above the Galactic plane. Distributions shown are the averages over 200 maps, where error bars indicate the corresponding standard deviations. Left to right panels show the results for the original NE2001 model, the modified NE2001 model (our model), and the plane-parallel model, respectively. Symbols denote the pulsar observations at high galactic latitude (40 - 90 degrees) used by \citet{gmcm08}; filled and open circles indicate observations with distance determination from trigonometric parallaxes and from associations with globular clusters, respectively.} \label{figure11}
}
\end{figure*}

\begin{figure*}[tp]
\figurenum{12}
\epsscale{0.9}
\plotone{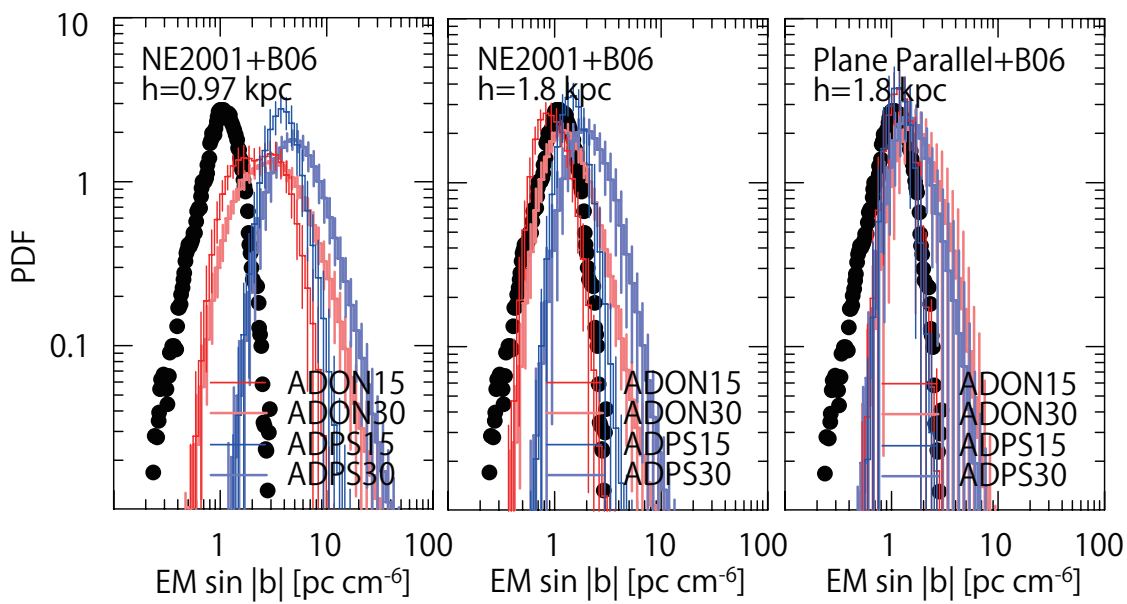}
\caption{
{Probability distribution functions (PDFs) of EM over a $60^\circ\times 60^\circ$ FOV toward the north and south Galactic poles. Shown PDFs are the averages for 200 maps, where error bars indicate the standard deviation. Left to right panels show the results for the original NE2001 model, the modified NE2001 model (our model), and the plane-parallel model. We adopted the volume filling factor introduced by \citet{ber06}. Filled circles are the WHAM observations at high (60 - 90 degree) galactic latitude \citep{hil08}.} \label{figure12}
}
\end{figure*}

\begin{figure*}[tp]
\figurenum{13}
\epsscale{0.9}
\plotone{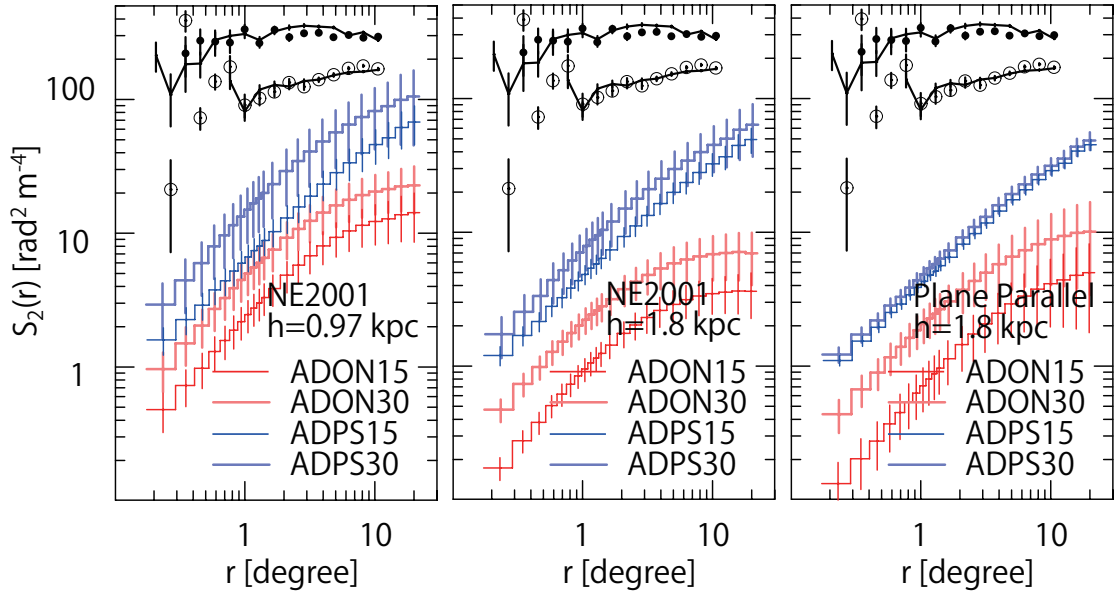}
\caption{
{Second-order structure functions (SFs) of simulated RMs over a $30^\circ\times 30^\circ$ FOV. Shown SFs are the averages for 200 maps, where error bars indicate the standard deviation. Left to right panels show the results for the original NE2001 model, the modified NE2001 model (our model), and the plane-parallel model. Open circles and black thick lines are the observed second-order SFs toward the NGP, and filled circles and black thin lines are those toward the SGP (\citet[][circles]{mao10} and \citet[][lines]{sts11}).} \label{figure13}
}
\end{figure*}

Reversals in the sign of RM across the Galactic plane and across the Galactic center \citep[e.g.,][]{tss09} suggest the existence of a halo toroidal field. Global three-dimensional MHD simulations of gas disks \citep{nmm06} have indicated the azimuthal magnetic field component in the halo as a result of the buoyant escape of the azimuthal magnetic flux from the disk. The halo toroidal (azimuthal) field can be expressed as
\begin{eqnarray}
\left\{
\renewcommand{\arraystretch}{1.5}
\begin{array}{ll}
B_{{\rm t},R}(R,\Theta,z) & = 0,\\
B_{{\rm t},\Theta}(R,\Theta,z) & =
\displaystyle{
\frac{{\rm sign}(z)^vB_{\rm t0}}{1+\left( \frac{|z|-z_{\rm t0}}
{z_{\rm t1}} \right)^2}\frac{R}{R_{\rm t0}}
\exp \left( -\frac{R-R_{\rm t0}}{R_{\rm t0}}\right),}\\
B_{{\rm t},z}(R,\Theta,z) & = 0,
\end{array}
\renewcommand{\arraystretch}{1.0}
\right.
\end{eqnarray}
where ${\rm sign}(z)$ is the sign of $z$, and $v$ introduces the parity of the toroidal field configuration; $v=1$ if we consider the asymmetries in longitude and latitude relative to the Galactic plane and the center, respectively (dipole), or $v=2$ if we consider the axisymmetric configuration without reversals relative to the Galactic plane (quadrupole). We adopt $B_{\rm t0}=2~{\rm \mu G}$, $z_{\rm t0}=1.5~{\rm kpc}$, $R_{\rm t0}=4~{\rm kpc}$, and $z_{\rm t1}=0.2~{\rm kpc}$ for $|z|<z_{\rm t0}$ and $z_{\rm t1}=4.0~{\rm kpc}$ otherwise. We note that \cite{ps03} adopted smaller $z_{\rm t1}$ for high $z$ (and also similarly small $z_{\rm t2}$ in their model), which results in $B_{\rm t}$ of order $0.1\ \mu{\rm G}$ at $z\sim 3~{\rm kpc}$. We adopt the parameters based on \cite{sr10}, which gives $\sim 1\ \mu{\rm G}$ at high $z$ (see also Figure \ref{figure3}). Such a strong halo field at high $z$ is also motivated by recent studies of halo magnetic fields \citep[see][]{mao12, jf12}.

Nonthermal filaments observed near the Galactic center imply the existence of a central, vertical magnetic field \citep{han09}, which is also predicted from global three-dimensional MHD simulations of gas disks \citep[e.g.,][]{mac09}. The vertical field can be due to the Galactic-center poloidal (dipole) field; a strong dipole field would be observed as the vertical magnetic field in the Earth vicinity. The dipole field can be expressed as 
\begin{eqnarray}
\left\{
\renewcommand{\arraystretch}{2.2}
\begin{array}{ll}
B_{{\rm p},R}(R,\Theta,z) & = \displaystyle -\frac{\mu_{\rm p}}{(R^2+z^2)^{3/2}}
\frac{3Rz}{R^2+z^2},\\
B_{{\rm p},\Theta}(R,\Theta,z) & = 0,\\
B_{{\rm p},z}(R,\Theta,z) & = \displaystyle \frac{\mu_{\rm p}}{(R^2+z^2)^{3/2}}
\left(\frac{3z^2}{R^2+z^2} -1\right).
\end{array}
\renewcommand{\arraystretch}{1.0}
\right.
\end{eqnarray}
Here, we set $\mu_{\rm p}=180~{\rm \mu G \cdot kpc^3}$ \citep{gkss10} to make the vertical component of $\sim 0.3~{\rm \mu G}$ from the poloidal field. The vertical field could be due to the out-of-plane X-field recently studied by \cite{jf12}. This model is motivated by the X-shaped field structure seen in radio observations of external, edge-on galaxies \citep{beck09b, krause09}. We define the elevation angle of magnetic field, $\eta(R, \Theta, z)$, with respect to the Galactic mid-plane. Then, the X-field can be expressed as
\begin{eqnarray}
\left\{
\renewcommand{\arraystretch}{2.2}
\begin{array}{ll}
B_{{\rm x},R}(R,\Theta,z) & = \displaystyle {\rm sign}(z) B_{\rm x0} \exp\left( -\frac{R_{\rm p}}{R_{\rm x0}}\right) \left(\frac{R_{\rm p}}{R}\right)^w\cos\eta,\\
B_{{\rm x},\Theta}(R,\Theta,z) & = 0,\\
B_{{\rm x},z}(R,\Theta,z) & = \displaystyle B_{\rm x0} \exp\left( -\frac{R_{\rm p}}{R_{\rm x0}}\right) \left(\frac{R_{\rm p}}{R}\right)^w\sin\eta.
\end{array}
\renewcommand{\arraystretch}{1.0}
\right.
\end{eqnarray}
We take the field at $R_{\rm p}> R_{\rm xc}$ to have a constant elevation angle, $\eta_{\rm x0}$, with respect to the mid-plane, where $R_{\rm p}$ is the radius at which the field line passing through $(R, \Theta, z)$ crosses the mid-plane. This means that if $R-|z|/\tan\eta_{\rm x0} > R_{\rm x0}$, then $R_{\rm p}=R-|z|/\tan\eta_{\rm x0}$, $\eta=\eta_{\rm x0}$, and $w=1$. Otherwise, $R_{\rm p}=RR_{\rm xc}/(R_{\rm xc}+|z|/\tan\eta_{\rm x0})$ $\eta=\tan^{-1}\{|z|/(R-R_{\rm p})\}$ and $w=2$. We adopt $B_{\rm x0}=4.6~{\rm \mu G}$, $\eta_{\rm x0}=49^\circ$, $R_{\rm xc}=4.8~{\rm kpc}$, and $R_{\rm x0}=2.9~{\rm kpc}$.

\acknowledgments

We thank the referee for constructive comments. We also thank P. L. Biermann for comments on the manuscript, M. Machida and X. Sun for discussions on the Galactic magnetic field model, S. A. Mao for providing the data used in Figure 6, and J. Stil for providing the data used in Figures 9 and 12. T.A. acknowledges the supports of the Korea Research Council of Fundamental Science and Technology (KRCF) and the Japan Society for the Promotion of Science (JSPS). D.R. acknowledges the support of the National Research Foundation (NRF) of Korea through grant 2007-0093860. J.K. acknowledges the support by the International Research \& Development Program of the National Research Foundation (NRF) funded by the Ministry of Education, Science and Technology (MEST) of Korea (grant K20903001740-11B1300-02610). B.M.G. acknowledges the support of the Australian Research Council through grant FL100100114.

\clearpage

\begin{deluxetable}{lcccc}
\tablenum{1}
\tabletypesize{\footnotesize}
\tablewidth{0pt}
\tablecaption{Models, regular field direction, Mach number, and $\beta_0$
\label{table1}
}
\tablehead{
\colhead{Model\tablenotemark{a}} 
& \colhead{Block 1} & \colhead{Block 2} & \colhead{Block 3} & \colhead{Block 4}\\
& $(0.00\le |z|<1.25)$\tablenotemark{b} & $(1.25\le |z|<2.50)$\tablenotemark{b} & 
$(2.50\le |z|<5.00)$\tablenotemark{b} & $(5.00\le |z|<10.0)$\tablenotemark{b}\\
& $\textit{\textbf{B}}_0$, $M_{\rm rms}$, $\beta_0$
& $\textit{\textbf{B}}_0$, $M_{\rm rms}$, $\beta_0$
& $\textit{\textbf{B}}_0$, $M_{\rm rms}$, $\beta_0$
& $\textit{\textbf{B}}_0$, $M_{\rm rms}$, $\beta_0$
}
\startdata
ADON15 & $+x$, 1.0, 3.0 & $-x$, 1.0, 3.0 & $-x$, 0.5, 0.1 & $-x$, 0.5, 0.1 \\
ADOS15 & $+x$, 1.0, 1.0 & $+x$, 1.0, 0.1 & $+x$, 0.5, 0.1 & $+x$, 0.5, 0.1 \\
AQOS15 & $+x$, 1.0, 3.0 & $-x$, 1.0, 3.0 & $-x$, 0.5, 0.1 & $-x$, 0.5, 0.1 \\
ADPN15 & $+x$, 1.0, 1.0 & $-x$, 1.0, 1.0 & $-x$, 0.5, 0.1 & $-x$, 0.5, 0.1 \\
ADPS15 & $+x$, 1.0, 1.0 & $+x$, 1.0, 0.1 & $+x$, 0.5, 0.1 & $+x$, 0.5, 0.1 \\
AQPS15 & $+x$, 1.0, 1.0 & $-x$, 1.0, 1.0 & $-x$, 0.5, 0.1 & $-x$, 0.5, 0.1 \\
ADXN15 & $+x$, 1.0, 1.0 & $-x$, 1.0, 1.0 & $-x$, 0.5, 0.1 & $-x$, 0.5, 0.1 \\
ADXS15 & $+x$, 1.0, 0.1 & $+x$, 1.0, 0.1 & $+x$, 0.5, 0.1 & $+x$, 0.5, 0.1 \\
AQXS15 & $+x$, 1.0, 1.0 & $-x$, 1.0, 1.0 & $-x$, 0.5, 0.1 & $-x$, 0.5, 0.1 \\
BDON15 & $-x$, 1.0, 3.0 & $-x$, 1.0, 3.0 & $-x$, 0.5, 0.1 & $-x$, 0.5, 0.1 \\
BDOS15 & $+x$, 1.0, 1.0 & $+x$, 1.0, 0.1 & $+x$, 0.5, 0.1 & $+x$, 0.5, 0.1 \\
BQOS15 & $-x$, 1.0, 3.0 & $-x$, 1.0, 3.0 & $-x$, 0.5, 0.1 & $-x$, 0.5, 0.1 \\
BDPN15 & $-x$, 1.0, 1.0 & $-x$, 1.0, 1.0 & $-x$, 0.5, 0.1 & $-x$, 0.5, 0.1 \\
BDPS15 & $+x$, 1.0, 1.0 & $+x$, 1.0, 0.1 & $+x$, 0.5, 0.1 & $+x$, 0.5, 0.1 \\
BQPS15 & $-x$, 1.0, 1.0 & $-x$, 1.0, 1.0 & $-x$, 0.5, 0.1 & $-x$, 0.5, 0.1 \\
BDXN15 & $-x$, 1.0, 1.0 & $-x$, 1.0, 1.0 & $-x$, 0.5, 0.1 & $-x$, 0.5, 0.1 \\
BDXS15 & $+x$, 1.0, 1.0 & $+x$, 1.0, 0.1 & $+x$, 0.5, 0.1 & $+x$, 0.5, 0.1 \\
BQXS15 & $-x$, 1.0, 1.0 & $-x$, 1.0, 1.0 & $-x$, 0.5, 0.1 & $-x$, 0.5, 0.1 \\
ADON30 & $+x$, 2.0, 1.0 & $-x$, 2.0, 1.0 & $-x$, 1.0, 0.1 & $-x$, 0.5, 0.1 \\
ADOS30 & $+x$, 2.0, 1.0 & $+x$, 2.0, 0.1 & $+x$, 1.0, 0.1 & $+x$, 0.5, 0.1 \\
AQOS30 & $+x$, 2.0, 1.0 & $-x$, 2.0, 1.0 & $-x$, 1.0, 0.1 & $-x$, 0.5, 0.1 \\
ADPN30 & $+x$, 2.0, 1.0 & $-x$, 2.0, 1.0 & $-x$, 1.0, 0.1 & $-x$, 0.5, 0.1 \\
ADPS30 & $+x$, 2.0, 1.0 & $+x$, 2.0, 0.1 & $+x$, 1.0, 0.1 & $+x$, 0.5, 0.1 \\
AQPS30 & $+x$, 2.0, 1.0 & $-x$, 2.0, 1.0 & $-x$, 1.0, 0.1 & $-x$, 0.5, 0.1 \\
ADXN30 & $+x$, 2.0, 1.0 & $-x$, 2.0, 1.0 & $-x$, 1.0, 0.1 & $-x$, 0.5, 0.1 \\
ADXS30 & $+x$, 2.0, 0.1 & $+x$, 2.0, 0.1 & $+x$, 1.0, 0.1 & $+x$, 0.5, 0.1 \\
AQXS30 & $+x$, 2.0, 1.0 & $-x$, 2.0, 1.0 & $-x$, 1.0, 0.1 & $-x$, 0.5, 0.1 \\
BDON30 & $-x$, 2.0, 1.0 & $-x$, 2.0, 1.0 & $-x$, 1.0, 0.1 & $-x$, 0.5, 0.1 \\
BDOS30 & $+x$, 2.0, 1.0 & $+x$, 2.0, 0.1 & $+x$, 1.0, 0.1 & $+x$, 0.5, 0.1 \\
BQOS30 & $-x$, 2.0, 1.0 & $-x$, 2.0, 1.0 & $-x$, 1.0, 0.1 & $-x$, 0.5, 0.1 \\
BDPN30 & $-x$, 2.0, 1.0 & $-x$, 2.0, 1.0 & $-x$, 1.0, 0.1 & $-x$, 0.5, 0.1 \\
BDPS30 & $+x$, 2.0, 1.0 & $+x$, 2.0, 0.1 & $+x$, 1.0, 0.1 & $+x$, 0.5, 0.1\\
BQPS30 & $-x$, 2.0, 1.0 & $-x$, 2.0, 1.0 & $-x$, 1.0, 0.1 & $-x$, 0.5, 0.1 \\
BDXN30 & $-x$, 2.0, 1.0 & $-x$, 2.0, 1.0 & $-x$, 1.0, 0.1 & $-x$, 0.5, 0.1 \\
BDXS30 & $+x$, 2.0, 1.0 & $+x$, 2.0, 0.1 & $+x$, 1.0, 0.1 & $+x$, 0.5, 0.1 \\
BQXS30 & $-x$, 2.0, 1.0 & $-x$, 2.0, 1.0 & $-x$, 1.0, 0.1 & $-x$, 0.5, 0.1
\enddata
\tablenotetext{a}{
A: axi-symmetric spiral,
B: bi-symmetric spiral,
D: dipole toroidal,
Q: quadrupole toroidal,
O: no poloidal,
P: dipole poloidal,
X: X-field poloidal,
15: random field with rms flow speed 15~${\rm km/s}$,
30: random field with rms flow speed 30~${\rm km/s}$,
N: toward the NGP,
S: toward the SGP.
}
\tablenotetext{b}{
In units of kpc.
}
\end{deluxetable}

\clearpage

\begin{deluxetable}{llrr}
\tablenum{2}
\tabletypesize{\small}
\tablewidth{0pt}
\tablecaption{
Rms values of density and magnetic field strength in MHD turbulence simulations. \label{table3}
}
\tablehead{
\colhead{$M_{\rm rms}$} & \colhead{$\beta_0$} 
& \colhead{$\rho_{\rm rms}/\rho_0$} & \colhead{$B_{\rm rms}/B_0$}
}
\startdata
0.5
& 0.1 & 1.02702 & 1.00360 \\
& 1.0 & 1.03139 & 1.03748 \\
& 10. & 1.02755 & 1.34981 \\
1.0
& 0.1 & 1.09307 & 1.01275 \\
& 1.0 & 1.11417 & 1.14356 \\
& 3.0 & 1.10303 & 1.40630 \\
& 10. & 1.11177 & 2.09291 \\
2.0
& 0.1 & 1.29744 & 1.04641 \\
& 1.0 & 1.32886 & 1.45009 \\
& 10. & 1.32293 & 3.04786
\enddata
\end{deluxetable}

\clearpage

\begin{deluxetable}{lrrcrr}
\tablenum{3}
\tabletypesize{\small}
\tablewidth{0pt}
\tablecaption{
The average, $\mu$, and standard deviation, $\sigma$, of the RMs due to the regular components of the electron density and the GMF. \label{table2}
}
\tablehead{
\colhead{} & \multicolumn{2}{c}{$200~{\rm deg^2}$} & & \multicolumn{2}{c}{$900~{\rm deg^2}$} \\
\cline{2-3} \cline{5-6} \\
\colhead{Model} & 
\colhead{$\mu$\tablenotemark{a}} &
\colhead{$\sigma$\tablenotemark{a}} & &
\colhead{$\mu$\tablenotemark{a}} &
\colhead{$\sigma$\tablenotemark{a}}
}
\startdata
ADON & 0.00 & 0.18 && 0.06 & 0.44\\
ADOS & 0.02 & 1.92 && 0.03 & 4.21\\
AQOS & 0.02 & 0.39 && 0.03 & 0.90\\
ADPN & -4.76 & 0.45 && -4.83 & 0.87\\
ADPS & 5.58 & 1.95 && 5.66 & 4.22\\
AQPS & 5.58 & 0.36 && 5.65 & 0.78\\
ADXN & -4.12 & 0.13 && -4.12 & 0.29\\
ADXS & 4.69 & 1.97 && 4.70 & 4.28\\
AQXS & 4.69 & 0.51 && 4.70 & 1.11\\
BDON & 0.00 & 0.14 && 0.05 & 0.35\\
BDOS & 0.02 & 1.87 && 0.03 & 4.08\\
BQOS & 0.02 & 0.32 && 0.03 & 0.75\\
BDPN & -4.76 & 0.41 && -4.83 & 0.80\\
BDPS & 5.58 & 1.90 && 5.65 & 4.09\\
BQPS & 5.58 & 0.31 && 5.65 & 0.66\\
BDXN & -4.12 & 0.10 && -4.12 & 0.24\\
BDXS & 4.69 & 1.91 && 4.70 & 4.14\\
BQXS & 4.68 & 0.44 && 4.70 & 0.96
\enddata
\tablenotetext{a}{
In units of ${\rm rad~m^{-2}}$.
}
\end{deluxetable}

\clearpage

\clearpage

\clearpage

\clearpage

\clearpage

\clearpage

\clearpage

\clearpage

\clearpage

\clearpage

\clearpage

\clearpage

\clearpage

\end{document}